\documentclass{nature}

\usepackage{graphicx} % Required for inserting images

\usepackage{amsmath}
\usepackage{amssymb}
\usepackage{amsfonts}
\usepackage{mathbbol}
\usepackage{physics}
\usepackage{hyperref}
\usepackage{threeparttable}

\usepackage{siunitx}
\usepackage[symbol]{footmisc}

\spacing{1.0}
\newcommand\aj{{Astron.\ J.}}% 
\newcommand\araa{{Annu.\ Rev.\ Astron.\ Astr.}}% 
\newcommand\apj{{Astrophys.\ J.}}% 
\newcommand\apjl{{Astrophys.\ J.\ Lett.}}% 
\newcommand\apjs{{Astrophys.\ J.\ Supp.}}% 
\newcommand\aap{{Astron.\ Astrophys.}}% 
\newcommand\mnras{{Mon.\ Not.\ R.\ Astron.\ Soc.}}% 
\newcommand\pasp{{Publ.\ Astron.\ Soc.\ Pac.}}% 
\newcommand\nat{{Nature}}% 
\newcommand\rmxaa{{Rev.\ Mex.\ Astron.\ Astrofis.}}
\newcommand\kms{\ensuremath{{\rm km}~{\rm s}^{-1}}}
\newcommand\neviwave{[Ne\,\textsc{vi}] $\lambda$7.652 $\mu$m}
\newcommand\nevi{[Ne\,\textsc{vi}]}
\newcommand\neiiwave{[Ne\,\textsc{ii}] $\lambda$12.813 $\mu$m}
\newcommand\neii{[Ne\,\textsc{ii}]}
\newcommand\nevwave{[Ne\,\textsc{v}] $\lambda$14.322 $\mu$m}
\newcommand\nev{[Ne\,\textsc{v}]}
\newcommand\angstrom{\ensuremath{\mathring{\rm A}}}
\newcommand\ergs{\ensuremath{{\rm erg}~{\rm s}^{-1}}}
\newcommand\ergscm{\ensuremath{{\rm erg}~{\rm s}^{-1}~{\rm cm}^{-2}}}
\newcommand\msun{\ensuremath{M_\odot}}
\newcommand\msunyr{\ensuremath{M_\odot~{\rm yr}^{-1}}}

\newcommand\angst{\ensuremath{\mathring{\rm A}}}
\newcommand\oviwave{O\,\textsc{vi} $\lambda\lambda$1032,1038\angst}
\newcommand\ovi{O\,\textsc{vi}}
\newcommand\rev{}

\title{Directly \rev{Imaging} the Cooling Flow in the Phoenix Cluster}
\author{Michael Reefe$^1$, Michael McDonald$^1$, Marios Chatzikos$^2$, Jerome Seebeck$^3$, Richard Mushotzky$^3$, Sylvain Veilleux$^3$, Steven W. Allen$^{4,5,6}$, Matthew Bayliss$^7$, Michael Calzadilla$^1$, Rebecca Canning$^8$, Benjamin Floyd$^9$, Massimo Gaspari$^{10}$, Julie Hlavacek-Larrondo$^{11}$, Brian McNamara$^{12}$, Helen Russell$^{13}$, Keren Sharon$^{14}$, and Taweewat Somboonpanyakul$^{4,5}$}
\date{June 2024}

\begin{document}

\maketitle
\begin{affiliations}
\item Kavli Institute for Astrophsyics \& Space Research, Massachusetts Institute of Technology, Cam-bridge, MA 02139, USA
\item Department of Physics \& Astronomy, University of Kentucky, Lexington, KY 40506, USA
\item Department of Astronomy \& Joint Space-Science Institute, University of Maryland, College Park, MD 20742, USA
\item Kavli Institute for Particle Astrophysics and Cosmology, Stanford University, 452 Lomita Mall, Stanford, CA 94305, USA
\item Department of Physics, Stanford University, 382 Via Pueblo Mall, Stanford, CA 94305, USA
\item SLAC National Accelerator Laboratory, 2575 Sand Hill Road, Menlo Park, CA 94025, USA
\item Department of Physics, University of Cincinnati, Cincinnati, OH 45221, USA
\item Institute of Cosmology \& Gravitation, University of Portsmouth, Dennis Sciama Building, PO1 3FX, UK
\item Department of Physics and Astronomy, University of Missouri-Kansas City, Flarsheim Hall, 5110 Rockhill Road, Kansas City, MO 64110, USA
\item Department of Physics, Informatics and Mathematics, University of Modena and Reggio Emilia, 41125 Modena, Italy
\item D\'epartement de Physique, Universit\'e de Montr\'eal, C.P. 6 128, Succ. Centre-Ville, Montr\'eal, Qu\'ebec H3C 3J7, Canada
\item Department of Physics and Astronomy, University of Waterloo, Waterloo, ON N2L 3G1, Canada
\item School of Physics \& Astronomy, University of Nottingham, University Park, Nottingham NG7 2RD, UK
\item Department of Astronomy, University of Michigan, 1085 S. University Ave, Ann Arbor, MI 48109, USA
\end{affiliations}

% Gas phases:
% \cite{2001MNRAS.328..762E, 2002MNRAS.337...49E, 2003A&A...412..657S, 2008A&A...483..793S, 1985ApJS...59..447H, 1987MNRAS.224...75J, 1989ApJ...338...48H, 1999MNRAS.306..857C, 2007MNRAS.379..100E, 2007MNRAS.380...33H, 2010ApJ...721.1262M, 2011ApJ...731...33M, 2001ApJ...553L.125B, 2006ApJ...642..746B, 2001ApJ...560..187O, 2014ApJ...791L..30M}.  

% Star formation rates:
% \cite{1994ARA&A..32..277F, 1989AJ.....98.2018M, 1995MNRAS.276..947A, 1999MNRAS.306..857C, 2005ApJ...635L...9H, 2007MNRAS.379..100E, 2007MNRAS.380...33H, 2008ApJ...681.1035O, 2010ApJ...721.1262M, 2018ApJ...858...45M, 2012ApJS..199...23H, 2012ApJ...747...29R, 2015ApJ...805..177D, 2015MNRAS.450.2564M, 2016A&A...595A.123M}.  

\begin{abstract}
% Clusters of galaxies are comprised of hundreds of galaxies, typically with a single massive galaxy at the center of the cluster dominating the local gravitational potential.
In the centers of many galaxy clusters, the hot ($\sim$10$^7$ K) intracluster medium (ICM)
%---a diffuse plasma that fills the space between galaxies and outweighs the stars by a factor of ten---
can become dense enough that it should cool on short timescales\cite{1984Natur.310..733F, 1994ARA&A..32..277F}.
%to low temperatures within the age of the universe
However, the low measured star formation rates in massive central galaxies\cite{1989AJ.....98.2018M, 1995MNRAS.276..947A, 2005ApJ...635L...9H, 2018ApJ...858...45M} and absence of soft X-ray lines from cooling gas\cite{1988ASIC..229...63C, 2001ApJ...557..546D, 2003ApJ...590..207P} suggest that most of this gas never cools -- this is known as the ``cooling flow problem.'' The latest observations suggest that black hole jets are maintaining the vast majority of gas at high temperatures\cite{2007ARA&A..45..117M, 2012ARA&A..50..455F, 2015ApJ...805...35H, 2011MNRAS.411..349G, 2015ApJ...811..108P, 2017ApJ...847..106L, 2019ApJ...871....6Y}. A cooling flow has yet to be fully mapped through all gas phases in any galaxy cluster. Here, we present new observations of the Phoenix cluster\cite{2012Natur.488..349M} using the \textit{James Webb Space Telescope} to map the 
%several high-ionization lines in the infrared on $>$20 kpc scales, including the 
\neviwave\ emission line, allowing us to probe gas at 10$^{5.5}$ K on large scales. These data show extended \nevi\, emission  cospatial with \rev{(i)} the cooling peak in the ICM, \rev{(ii) the coolest gas phases}, and \rev{(iii) sites of active star formation}. \rev{Taken together, these imply a recent episode of rapid cooling, causing a short-lived spike in the cooling rate which we estimate to be} 5,000--23,000 \msunyr. %\rev{This cooling is occurring} behind a buoyantly rising X-ray bubble\rev{, perhaps seeded by the turbulent wake or by uplift of low-entropy gas.}
These data provide the first \rev{large-scale map} of \rev{gas at temperatures between 10$^5$--10$^6$ K} in a \rev{cluster core}, and highlight the critical role that black hole feedback plays in not only regulating but also promoting cooling\cite{2016ApJ...830...79M}.
\end{abstract}

%%%%%%%%%%%%%%%%%%%%%%%%%%%%%%%% MAIN TEXT %%%%%%%%%%%%%%%%%%%%%%%%%%%

In many galaxy clusters, the hot intracluster medium (ICM) in the inner $\sim$100\,kpc is dense and cool enough that it should rapidly cool and fuel a massive starburst in the central galaxy on timescales much shorter than the age of the universe\cite{1984Natur.310..733F}.  Such ``cooling flows'' are largely absent, with observed star formation rates (SFRs) in central cluster galaxies rarely exceeding 1--10 \msunyr [\citenum{1994ARA&A..32..277F, 1989AJ.....98.2018M, 1995MNRAS.276..947A, 2005ApJ...635L...9H, 2018ApJ...858...45M}].  This implies the presence of energy sources that suppress the cooling, with the most likely being feedback from an active galactic nucleus (AGN). X-ray observations suggest this is achieved by bipolar radio jets emitted from the AGN, which inject mechanical energy and drive turbulence in the surrounding ICM and inflate large buoyant cavities or ``bubbles'' of plasma that can transport low-entropy gas outwards\cite{2007ARA&A..45..117M, 2008A&A...477L..33R, 2012ARA&A..50..455F, 2015ApJ...805...35H, 2016ApJ...830...79M}. Simulations modeling the underlying physics show that this feedback can be recurrently induced via feeding of the supermassive black hole, e.g. through thermally unstable cooling and cold gas precipitation, intrinsically linking the processes of feeding and feedback\cite{2011MNRAS.411..349G, 2015ApJ...811..108P, 2017ApJ...847..106L, 2019ApJ...871....6Y}. In this context, the Phoenix Cluster\cite{2012Natur.488..349M, 2019ApJ...885...63M} \rev{($z=0.597$)} is an outlier, as it seems to follow all of the original predictions for an uninhibited cooling flow.  It has an enormous starburst\cite{2012Natur.488..349M}, a massive cold gas reservoir\cite{2017ApJ...836..130R}, a rapidly accreting supermassive black hole powering an AGN in the central galaxy\cite{2019ApJ...885...63M}, a power-law entropy profile all the way into the core\cite{2019ApJ...885...63M}, and a central cooling time comparable to the freefall time and significantly shorter than the turbulent eddy timescale (which measures the time for a turbulent vortex to oscillate and produce density fluctuations in the surrounding medium)\cite{2018ApJ...854..167G, 2019ApJ...885...63M}. 

Previous studies attempting to directly detect \rev{the Phoenix cluster's} cooling flow have yielded mixed results. \rev{Coronal emission lines, which probe gas at intermediate temperatures (10$^5$--10$^6$ K), provide a promising avenue for directly tracing cooling flows.} Observations of the \oviwave\, coronal line doublet in the Phoenix Cluster with the \textit{Hubble Space Telescope's} Cosmic Origins Spectrograph (\textit{HST}-COS) imply massive amounts of cooling (55,000 \msunyr) when assuming 100\% of the observed flux is due to cooling\cite{2015ApJ...811..111M}. But these observations lack spatial information, making the source of the \ovi\, emission ambiguous, and require a large, highly-uncertain correction for intrinsic extinction in the far-ultraviolet (UV).  Observations of soft X-ray lines with the \textit{X-ray Multi-Mirror Mission's} Reflection Grating Spectrometer (\textit{XMM}-RGS), on the other hand, find a cooling rate of $350^{+250}_{-200}$ \msunyr\, when fitting a combined isothermal gas and cooling flow model, which can barely sustain the 500--800 \msunyr\, starburst\cite{2018MNRAS.480.4113P}.  Cooling rates measured with these methods often disagree---\textit{Far Ultraviolet Spectroscopic Explorer} (\textit{FUSE}) observations of Abell 1795, Abell 2597, and Perseus all find \ovi\, cooling rates higher than the X-ray rates\rev{\cite{2001ApJ...560..187O,2006ApJ...642..746B}}. Like the \ovi\ line in the far-UV, soft X-rays may be obscured by cool, dusty clouds closely interleaved with the hot gas, causing the observed X-ray cooling rates to under-predict the amount of ``hidden'' cooling gas\cite{2022MNRAS.515.3336F}.
\rev{
Alternatively, the lack of soft X-ray emission may be explained if coronal emission is produced by the heating and mixing of ambient cold gas with the hot atmosphere, rather than the cooling of the hot atmosphere itself.  Observations of the C\,\textsc{iv} $\lambda$1549$\angstrom$ line in the Virgo cluster\cite{2012ApJ...750L...5S, 2016MNRAS.459.2806A} and the [Fe\,\textsc{x}] $\lambda$6374$\angstrom$ coronal line in the Centaurus cluster\cite{2011MNRAS.411..411C, 2015MNRAS.446.1234C} both seem to favor heating via thermal conduction or mixing over pure cooling flow models.}%  Unlike in the UV, optical coronal lines have less intrinsic dust extinction, but are instead complicated by their low emissivities, which can be completely lost in the noise of the underlying stellar continuum.  

In this work, we present new integral field unit (IFU) observations of the \neviwave\, coronal line, which probes a similar temperature range as \ovi\, ($\sim$10$^{5.5}$K) but in a wavelength regime with effectively no extinction \rev{and no stellar continuum}, using \textit{JWST}'s Mid-Infrared Instrument in the Medium Resolution Spectroscopy mode (MIRI/MRS). 
%At the redshift of the Phoenix Cluster ($z=0.597$), this line is at an observed wavelength of 12.220 $\mu$m. 
\rev{W}e develop a full mid-infrared (MIR) spectral model to fit the continuum and emission lines in each ``spaxel'' (\rev{spatial} pixel) of the IFU data,
% The continuum model includes thermal emission from dust at various temperatures, a broad silicate absorption feature at 9.7 $\mu$m, and broad emission features from Polycyclic Aromatic Hydrocarbons (PAHs). 
% Additionally, we separate the flux originating from the host galaxy and the bright AGN by applying a model of the MIRI/MRS point-spread function (PSF) combined with the spectrum of the central, brightest spaxel. 
performing a simultaneous spectral and spatial deconvolution of the central, dust-obscured QSO and the larger-scale emission. A map of the QSO-subtracted \nevi\, flux is shown in the left panel of Figure \ref{fig:overlay_city}. 
\rev{This map represents a dramatic improvement on previous observations of extended coronal emission\cite{2011MNRAS.411..411C,2006ApJ...642..746B}, which have detected 10$^{5.5}$\,K gas in only a handful of spaxels -- in Phoenix, we clearly detect \nevi\ emission in hundreds of spaxels over tens of kiloparsecs, allowing us to see coherent structure in the coronal phase for the first time.}
%\rev{This map represents an order of magnitude improvement on previous observations of extended coronal emission\cite{2011MNRAS.411..411C,2006ApJ...642..746B}, with clear detections in hundreds of spaxels over tens of kiloparsecs, allowing us to see coherent structure in the coronal phase for the first time.}
This map clearly shows two clouds of $\sim$10$^{5.5}$\,K gas to the north of the nucleus, cospatial with the minimum entropy of the hot atmosphere\cite{2019ApJ...885...63M}, the low- and high-ionization optical emission lines\cite{2014ApJ...784...18M}, the cold molecular gas\cite{2017ApJ...836..130R}, and a region of enhanced star formation (Figure \ref{fig:overlay_city}). 
% This is the first time that all phases of a cooling flow have been directly imaged.  
A one-dimensional radial intensity profile reveals that the \nevi\ emission is consistent with a large-scale diffuse nebula that is being photoionized by the central QSO and a localized cooling region with a size $\leqslant$5\,kpc at a distance of $\sim$10\,kpc from the nucleus (Figure \ref{fig:radialflux}). In principle, this bump in the radial profile could be caused by a local increase in the gas density $(n_e)$. However, we detect a similar bump in the \nevi/\nev\, ratio, which is insensitive to $n_e$, in the same region. Taken together, this implies a local change in the dominant ionization mechanism, such as from AGN photoionization to cooling. The presence of a cloud with an effective temperature $\sim$30$\times$ less than the lowest temperature seen in the X-ray-emitting ICM and $\sim$30\rev{$\times$} higher than the optical line emitting gas lying directly beneath the buoyantly-rising X-ray bubble could be related to either uplifting of low entropy gas from the core\cite{2008A&A...477L..33R, 2010MNRAS.406.2023P, 2016ApJ...830...79M} and/or rapid cooling in situ driven by turbulence\cite{2012ApJ...746...94G}, both of which can be explained via the AGN feedback model. In the latter case, turbulent motions result from compressions, rarefactions, and stretching in the surrounding medium, which can be induced by AGN jets, winds, and shocks from the bubble\cite{2020MNRAS.498.4983W}. 
%
%Turbulent motions result from compressions, rarefactions, and stretching in the surrounding medium, which can be induced by AGN jets, winds, and shocks from the bubble\cite{2020MNRAS.498.4983W}. The gas undergoes thermally unstable cooling and falls back down onto the nucleus through cold gas precipitation and accretion, fueling subsequent episodes of AGN feedback and causing the cycle to repeat. 
% The temperature of this coronal gas lies near the peak of the cooling curve\cite{1993ApJS...88..253S}, indicating that it is cooling rapidly relative to the hot atmosphere and will likely not be able to sustain cooling at this rate for long.

%We extract a MIR spectrum from a region to the north of the nucleus (marked by the white ellipse in Figure \ref{fig:overlay_city}) to further analyze the ionization state of the gas.  
%The \nevi\, intensity in this region is a few times higher than the fiducial $r^{-2}$ model (Figure \ref{fig:radialflux}), so we expect contributions from both a central ionizing source and some additional source of \nevi\, flux (assumed to be cooling).  
\rev{The total \nevi\, flux within this cloud north of the nucleus} (marked by the white ellipse in Figure \ref{fig:overlay_city}) is $4.9^{+0.5}_{-0.7} \times 10^{-16}$ \ergscm, which alone is enough to imply an unphysical cooling rate $>100,000~\msunyr$, motivating a more complex 2-component model.  We model the MIR spectrum using \textsc{Cloudy}\cite{1998PASP..110..761F, 2023RMxAA..59..327C} assuming that the emission is due to a combination of a central ionizing source and a cooling parcel of gas. The cooling gas starts with an initial density $n_e = 0.42$ cm$^{-3}$, temperature $kT = 2.3$ keV, and metallicity $Z = 0.55Z_\odot$, based on X-ray data at the location of the extended \nevi\ gas\cite{2019ApJ...885...63M}, and it cools until it reaches ISM temperatures ($10^4$\,K). Metal abundances are measured from high resolution X-ray spectroscopy of the central region of the cluster\cite{2018MNRAS.480.4113P} that have been re-computed for this work. While cooling, the gas is exposed to radiation from the central AGN.  While the central AGN in Phoenix is highly obscured along our line of sight, it is likely unobscured, or significantly less obscured, from the perspective of the cooling gas, as this gas lies along the jet axis where the AGN will not be obscured by the dusty torus.  We adopt a generic SED template constructed from a recent sample of hyperluminous quasars\cite{2023A&A...671A..34S} and rescale it to a bolometric luminosity of $3 \times 10^{47}$ \ergs, matching that of Phoenix\cite{2012Natur.488..349M}.  \rev{Cosmic rays are also included, with a background ionization rate of $2 \times 10^{-16}$ s$^{-1}$ [\citenum{2007ApJ...671.1736I}].}

The net effect of the AGN radiation is to lower the inferred cooling rates of the lines: Photoionization injects energy into the gas and increases the population of ions that contribute to the line luminosities, which increases the inferred luminosity per unit mass of cooling gas.  The sound-crossing time of the low-entropy gas of $\sim$10 Myr is comparable to the cooling time in the same region, both shorter than the cloud collapse time of $\sim$75 Myr, so the cooling is likely to proceed at some intermediate rate between the isobaric and isochoric predictions\cite{2015MNRAS.451L..60G}. Allowing for a \rev{combination} of these two modes and factoring in the photoionization from the central AGN, we obtain a nominal cooling rate for the \nevi\, line of 25,000 $\pm$ 5,000 \msunyr, where this uncertainty is purely statistical. When we consider all high-ionization lines in the spectrum (\nevi, \nev, [Fe\,\textsc{viii}], [Fe\,\textsc{vii}], [Mg\,\textsc{vii}], and O\,\textsc{vi}), the average cooling rate is 10,000 $\pm$ \rev{5,000} \msunyr\, (the uncertainty listed here is the median absolute deviation of the individual cooling rates).  Indeed, the cooling rates of all the IR lines are of a similar magnitude (Figure \ref{fig:nev_nevi_ratio}\rev{, left}), indicating that our model describes the overall spectrum reasonably well across the wide variety of ionization states of the IR lines (see Extended Data Table \ref{tab:lines}). The O\,\textsc{vi} line is an outlier, most likely due to uncertainties in the UV extinction correction. We explor\rev{e} the effects of varying the bolometric luminosity of the quasar and the hot phase abundance of neon, which (aside from the measured line fluxes) are the dominant drivers of the inferred cooling rates. We perfor\rev{m} additional simulations with these quantities higher/lower by a factor of two and f\rev{i}nd local scaling relations between the recovered cooling rates and the varied quantities for each line. Assuming the effects of the QSO luminosity and neon abundance combine multiplicatively, and combining measurements for all IR coronal lines, we find a plausible range of cooling rates %for the \nevi\, line between 12,000--52,000 \msunyr.  Similarly, for the average cooling rate of all the IR coronal lines, our plausible range is 
of 5,000--23,000 \msunyr\ through $\sim$10$^{5.5}$\,K.  

\rev{
We also consider a set of more complex models where gas from the hot $10^7$ K ICM first mixes with the warm $10^4$ K ISM in different proportions, cooling the hot atmosphere non-radiatively.  This mixed gas phase then cools radiatively down to $10^4$ K while illuminated by the AGN, in the same manner as our unmixed models.  The metal abundances in this model are treated as a mix of the ICM and ISM abundances, with the latter including the depletion of refractory elements onto dust grains\cite{2009ApJ...700.1299J}.  For more details, see the Supplementary Information.  We find that these models are able to produce more consistent cooling rates and line ratios than the unmixed models (Figure \ref{fig:nev_nevi_ratio}, right).  The average cooling rate is constrained to 15,000 $\pm$ 2,000 \msunyr, with a scatter 2.5$\times$ lower than the unmixed model.  The plausible range of cooling rates then becomes 7,000--36,000 \msunyr. We present these models as a promising alternative to the unmixed models and a showcase of how metal abundances in particular have a strong effect on the results, but we are cautious in claiming this scenario to be the underlying truth due to the additional systematic uncertainties introduced by the additional model parameters.  However, it is apparent from the results of both models that, whether the gas undergoes mixing or not, there is strong evidence for vast quantities of gas cooling through $10^{5.5}$ K in excess of 5,000 \msunyr.
}

Our most conservative estimate of the cooling rate (5,000--23,000 \msunyr) is significantly higher than the classical cooling rate inferred by the X-ray gas ($\sim$ 3,800 \msunyr). However, the latter is based on the bolometric X-ray luminosity and is averaged over several gigayears, whereas the emission line cooling measurements are nearly instantaneous. We propose that the high cooling rate, of order $10^4$ \msunyr\ at temperatures of $10^{5.5}$ K, is short-lived and not representative of the long-term cooling rate of this cluster\cite{2017MNRAS.466..677G, 2020MNRAS.495..594P}. This cooling spike is unlikely to last much longer than the sound-crossing time of the bubble ($\sim 10$ Myr), forming 0.5--2 $\times 10^{11}$ \msun\, of molecular gas. This is a few times higher than the observed $2.1 \pm 0.3 \times 10^{10}$ \msunyr [\citenum{2017ApJ...836..130R}], 
% reflective of the large systematic uncertainties not only in our analysis, but also in the cooling timescales and in the $\alpha_{\rm CO}$ conversion factor. Alternatively, 
\rev{which} could hint at large quantities of molecular gas being destroyed by feedback\cite{2022MNRAS.515.3336F}\rev{, or in the case of mixing, that some fraction of the gas is being prevented from cooling all the down to the molecular phase by repeatedly re-mixing with the hot ICM}. The \textsc{Cloudy} modeling is in excellent agreement with all of the observed line ratios, all of which are within a factor of a few of the model prediction. The only lines that disagree significantly are the low-ionization \neii\, line, which is likely boosted significantly by star formation and weak shocks, and O\,\textsc{vi}, again, likely due to a highly uncertain extinction correction. We show the results for the cooling rate\rev{s} in Figure \ref{fig:nev_nevi_ratio}\rev{, and the line ratios in Supplementary Information Figure \ref{fig:mixing_line_ratios}.}

Finally, we consider the velocity profile of the \nevi-emitting gas to further understand its origin. The coronal gas motions along our line of sight contain contributions from both bulk motions and turbulence imparted by the bubble, as evidenced by the large line widths (FWHMs $\simeq$ 1000 \kms). Despite this, the right panel of Figure \ref{fig:velocity_profile} demonstrates that the velocity profile exhibits coherent structure with radius.  We explore a toy model for the velocity profile of the cooling gas, considering the bulk motion imparted by the wake of a buoyantly rising bubble as it expands adiabatically in a stratified ICM\cite{2019ApJ...885...63M}. This model adequately describes the data at small radii out to $\sim$7\,kpc, at which point the gas may be slowing down due to its increased density relative to the ICM\cite{2003MNRAS.344L..48F, 2018MNRAS.479L..28G}. At radii larger than 7 kpc, the observed kinematics of the cooling gas are consistent with freefall. This suggests that the observed coronal emission may be attributed to low entropy gas that has been lifted from the core. For reference, a number of other velocity models are shown.  \rev{We also provide additional commentary on the morphology and kinematics of the coronal phase in comparison to the cooler gas phases in the Supplementary Information.

While the ideas behind cooling flows date back to the first high resolution images of galaxy clusters\cite{1984Natur.310..733F}, they have largely remained undetected. This may be because AGN feedback is highly effective in moderating cooling\cite{2007ARA&A..45..117M, 2012NJPh...14e5023M}, that episodes of pure cooling are common but short lived\cite{2017MNRAS.466..677G, 2020MNRAS.495..594P}, or because the majority of the cooling is obscured\cite{2022MNRAS.515.3336F, 2023MNRAS.521.1794F}. 
For the first time, we have mapped all phases of this cooling flow both in temperature and in space, providing the most complete picture of cooling to date. In the core of the Phoenix Cluster, we observe rapid cooling of the hot atmosphere, with traces of gas cascading through transitions in the UV, optical, infrared, and millimeter. Critically, this cooling gas is cospatial with the lowest entropy X-ray-emitting gas, lying directly behind a buoyantly-rising bubble in the hot atmosphere, suggesting that the gas is cooling in the turbulent wake behind the bubble, \rev{perhaps stimulated by first mixing with the cooler gas}.  If short-lived cooling episodes are common in the galaxy cluster population, providing the necessary fuel for ongoing AGN feedback, then Phoenix provides a unique window into this critically important, but rarely captured, process for understanding the formation of the most massive galaxies in the universe.  
% These data also serve as a window into what may be possible with next-generation X-ray IFUs (e.g. \textit{Athena}, \textit{Lynx}) and UV IFUs (e.g. \textit{Habitable Worlds Observatory}).

%It has long been postulated that the observed X-ray cavities in galaxy clusters are intrinsically linked to the cycle of cooling and feedback. These results, as the first spatially extended observations of coronal gas in a galaxy cluster, show direct evidence for this connection. The cooling rates, flux ratios, and kinematics of the coronal gas, in combination with its location just beneath the X-ray cavity, all paint a consistent picture of low entropy gas that has been uplifted in the wake of a buoyantly rising bubble and/or formed in situ through turbulence, where it can rapidly cool, condense, and rain back onto the core in the absence of direct AGN feedback. The cooling rates, while uncertain, are still 2 orders of magnitude higher than the soft X-ray cooling rates, hinting at strong X-ray absorption and lending credence to the idea of ``hidden'' cooling flows\cite{2022MNRAS.515.3336F} that may be present in other clusters as well. These data, which provide our first resolved view of a cooling flow, serve as a window into what may be possible with next-generation X-ray IFUs (e.g. Athena \& Lynx X-IFUs) or UV IFUs (e.g. \textit{Habitable Worlds Observatory}).

%%%%%%%%%%%%%%%%%%%%%%%%%%%%%%%% REFERENCES %%%%%%%%%%%%%%%%%%%%%%%%%%

\section*{References}

% \bibliography{main}{}
\bibliographystyle{nature}

%%%%%%%%%%%%%%%%%%%%%%%%%%%%%%%% FIGURES %%%%%%%%%%%%%%%%%%%%%%%%%%%%%

\clearpage

\begin{figure*}
\centering
\quad
\includegraphics[width=0.32\textwidth]{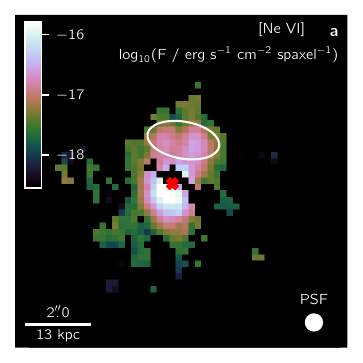}
\includegraphics[width=0.32\textwidth]{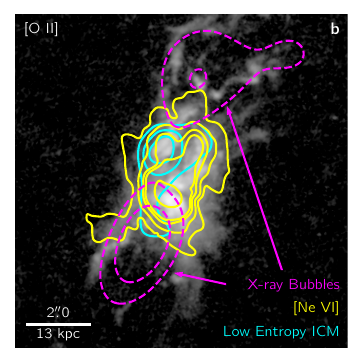}
\includegraphics[width=0.32\textwidth]{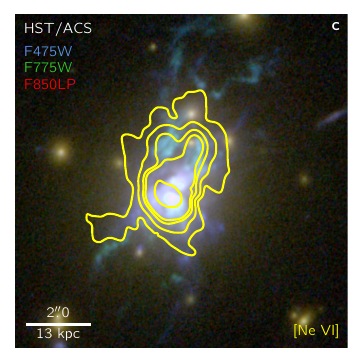}
\caption{\textbf{Maps of the \nevi-emitting coronal gas in the central galaxy of the Phoenix Cluster overlaid with the hotter and colder gas phases, and starlight.} (a) A 2D map of the \neviwave\ flux from channel 3 MIRI/MRS data. The flux is in $\log_{10}(F / \ergscm\,{\rm spaxel}^{-1})$, where a channel 3 spaxel is 0.04 arcsec$^2$. The white ellipse shows the aperture we use to capture the region of extended northern emission. A scale bar is provided in the lower-left, and the size of the PSF FWHM is in the lower-right. The red X marks the peak integrated brightness at $\ang[angle-symbol-over-decimal, angle-symbol-degree=^{\rm h}, angle-symbol-minute=^{\rm m}, angle-symbol-second=^{\rm s}]{23;44;43.9157}$, $\ang[angle-symbol-over-decimal]{-42;43;12.3972}$. Only spaxels where the line is detected with $S/N \geqslant 3$ are shown. (b) An [O\,\textsc{ii}] image of the central galaxy of the Phoenix Cluster is shown in greyscale using data from HST/ACS\cite{2019ApJ...885...63M}. The extended [O\,\textsc{ii}] emission is indicative of ongoing star formation.  The cyan contours show the entropy of the ICM, which decreases towards the center of the cluster and reaches a minimum in the northeast cloud. 
% The contours are at 7, 4.5, and 3 keV cm$^2$.  
The yellow contours show the flux of the \nevi\, line (as seen in panel a), which is cospatial with the brightest [O\,\textsc{ii}] and the lowest entropy gas. %These contours are at $-0.8$, 0, 0.6, 0.9, and 2.1$\sigma$ relative to the mean flux (in log space). 
The magenta dashed contours show the locations of the X-ray bubbles that are filled by radio jets, marking where the X-ray emission is $2\sigma$ and $4\sigma$ below the mean\cite{2019ApJ...885...63M}. (c) A 3-color RGB image of the central galaxy of the Phoenix Cluster is shown using data from HST/ACS in the F475W, F775W, and F850LP filters. Blue colors are indicative of young, actively forming stellar populations. The \nevi\, contours from the middle panel are shown here as well, highlighting the spatial coincidence between the blue star-forming filaments and the cooling gas.}
\label{fig:overlay_city}
\end{figure*}

\begin{figure*}
\centering
\includegraphics[width=\textwidth]{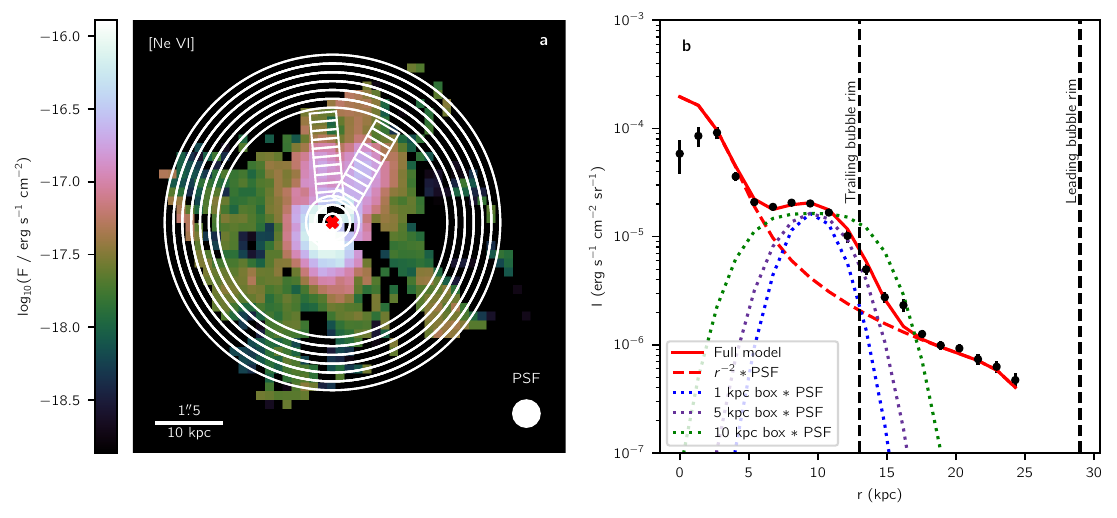}
\caption{\textbf{A raidal surface brightness profile of the northern \nevi\, emission in the central galaxy of the Phoenix Cluster.} (a) A 2D map of the \nevi\ flux. Here, only spaxels with $S/N \geqslant 1$ are shown. A series of rectangular apertures are shown in white that span from the nucleus out to 25 kpc in two directions aligned with the northern clouds of emission. The innermost and outermost apertures are circular annuli to improve the statistics. (b) A collapsed 1D intensity surface brightness profile is shown where measurements are taken by an average and standard deviation of the spaxels within each aperture. The profiles in the two directions shown in panel a are averaged to create the final 1D profile. The dashed red line shows an $r^{-2}$ power law convolved with the instrumental PSF, and the dotted lines show box functions with different widths in different colors, also convolved with the PSF. The solid red line is the combined model using the 5 kpc box width. The edges of the X-ray bubble are also annotated by vertical dashed lines at 13 and 29 kpc. The prominent bump in the profile that deviates from an $r^{-2}$ power law is coincident with the northern emission region beneath the bubble, and has a size of $\lesssim$5\,kpc.}% from $\sim 5$--13 kpc.}
\label{fig:radialflux}
\end{figure*}

\begin{figure*}
\centering
\quad
\includegraphics[width=0.48\textwidth]{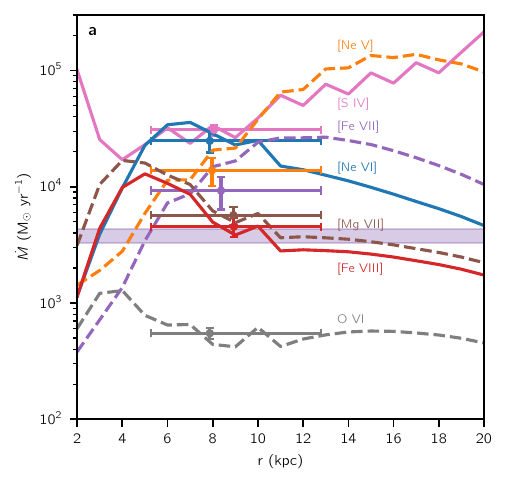}
\includegraphics[width=0.48\textwidth]{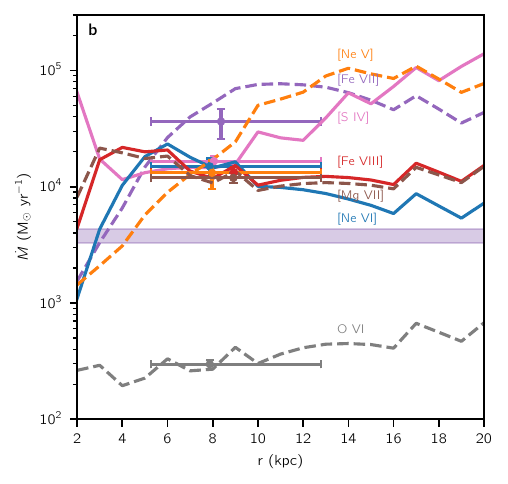}
\caption{\textbf{\textsc{Cloudy}-simulated cooling rates for a radiatively cooling parcel of gas that is being illuminated by a central AGN at various distances from the nucleus.} (a) The cooling rates $\dot{M}$ for each line are shown as a function of distance from the nucleus, assuming that all of the luminosity in the northern cloud is contained within a small $\Delta r$ at each radius. The lines are colored according to which emission line they correspond to. The intensity-weighted average cooling rates within the northern cloud are plotted with error bars---the $x$-errors show the extent of the northern cloud, the $x$-locations of the points themselves are the intensity-weighted central radii, and the $y$-errors show the statistical 1$\sigma$ uncertainties. The translucent purple band shows the inferred cooling rate from the X-ray emitting gas, which lies between the IR lines and O\,\textsc{vi}. (b) \rev{Same formatting as the left panel, but showing the results for our simulations that considered the initial gas conditions to be in a mixing layer between the hot and warm phases.}
}
\label{fig:nev_nevi_ratio}
\end{figure*}

\begin{figure*}
\centering
\includegraphics[width=\textwidth]{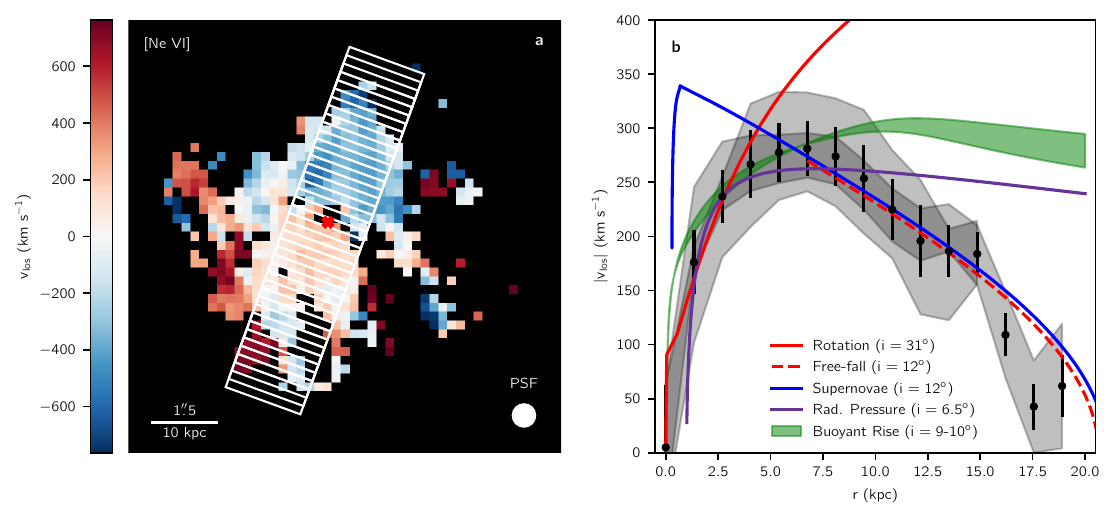}
\caption{\textbf{A north-south velocity profile of the \nevi-emitting gas in the central galaxy of the Phoenix Cluster, overlaid with various models.} (a) A 2D map of the median LOS velocity of \nevi. Here, only spaxels where \nevi\, is detected with $S/N \geqslant 2$ are shown. A series of rectangular apertures are shown in white that span the vertical height of the central galaxy (from $\sim$$-30$ to $+30$ kpc). (b) A collapsed 1D velocity profile is shown where measurements are taken by an SNR-weighted average and standard deviation of the spaxels within each aperture. The gray bands show the individual left- and right-side velocity profiles while the black data points show the average of these two profiles. A series of velocity models for pure rotation, ionized winds from supernovae, radiation pressure from the central AGN, and a buoyantly rising bubble are overplotted with varying inclination angles to best match the observed profile.}
\label{fig:velocity_profile}
\end{figure*}

\clearpage

%%%%%%%%%%%%%%%%%%%%%%%%%%%%%%%% METHODS %%%%%%%%%%%%%%%%%%%%%%%%%%%%%

\noindent{\large{\bf Methods}}

\section{Observations \& Data Reduction}
\label{sec:obs}

\subsection{Data Collection}
\label{sec:obs_collect}

\hfill\break
We obtained mid-infrared IFU observations of Phoenix A with JWST's MIRI/MRS instrument as a part of the cycle 1 General Observer's (GO) program ID 2439. The on-source observations were taken from UTC 2023 July 27 15:44:52 -- July 28 02:32:02, and background observations were taken from July 28 02:32:06 -- 04:48:42. The observations were taken in all channels/bands, which cover rest-frame wavelengths from 3.1--17.5 $\mu$m at $z=0.597$. The SHORT (A) band exposures were 6.95 hours, while the MEDIUM (B) and LONG (C) band exposures were 51.0 and 85.8 minutes, respectively. Note that the SHORT (A) exposure was longest because our primary goal was the detection of extended \nevi\, emission, which falls in that band. Background exposures were 51.0, 16.8, and 25.2 minutes, respectively. The on-source and background observations used the recommended 4-point and 2-point dither patterns, respectively. The field of view of each MIRI channel is shown in Extended Data Figure \ref{fig:miri_fov} over an optical image.

The HST/ACS data presented in Figure \ref{fig:overlay_city} and Extended Data Figure \ref{fig:miri_fov} was obtained from program ID 15315. Data was gathered in the broadband F475W, F775W, and F850LP filters for 2 orbits, 2 orbits, and 1 orbit respectively, and in the narrowband FR601N filter centered on [O\,\textsc{ii}] for 8 orbits. For details on this data and the data reduction, refer to [\citenum{2019ApJ...885...63M}]. 

The ICM entropy and X-ray cavity data presented in Figure \ref{fig:overlay_city} was modeled using data from \textit{Chandra} ACIS-I over a series of programs with IDs  13401, 16135, 16545, 19581, 19582, 19583, 20630, 20631, 20634, 20635, 20636, and 20797, totaling 551 ks of exposure time. Details on this data may also be found in [\citenum{2019ApJ...885...63M}].

\subsection{Data Reduction}
\label{sec:obs_reduce}

\hfill\break
We opt to reduce the data ourselves to allow the data reduction pipeline to be optimized for our observations, which are in a low surface brightness regime with deep exposures. We reduce the data using the JWST pipeline version 1.12.3 and CRDS context \texttt{jwst\_1140.pmap}. We follow the default pipeline options for most of the data reduction, with a few exceptions. We increase the sensitivity for flagging jumps from cosmic ray showers, and we perform an additional cleaning routine that finds warm pixels using the corresponding background exposures, marking them in both the background and science frames, following [\citenum{2023Natur.618..708S}]. We enable the 2D residual fringe correction step to suppress fringing on the individual spaxel level. Additionally, for the background subtraction step, we choose the option that directly subtracts a median-combined 2D background exposure pixel-by-pixel from the science exposures. This is in contrast to the default pipeline option that creates a 1D sigma-clipped background spectrum by combining all of the background spaxels at each wavelength, then subtracts this from the data in each spaxel.  The latter method has the advantage of introducing less noise into the background-subtracted data, but we have found that in our data this leaves residual detector artifacts that are better removed by performing the 2D background subtraction. We also produce cubes in the IFU detector aligned axes to allow direct comparison with the PSF. For more details on further refinements we perform on the data, see the Supplementary Information.

\section{Spectral Models}

\subsection{Model Decomposition}
\label{sec:method_spec_decomp}

\hfill\break
To model the spectra in each spaxel, we use a new fitting code written in the Julia programming language \cite{doi:10.1137/141000671}: Likelihood Optimization of gas Kinematics in IFUs (LOKI; \href{https://github.com/Michael-Reefe/Loki.jl}{https://github.com/Michael-Reefe/Loki.jl}). The total model can be summarized as:

\begin{align}
\label{eq:model}
\begin{split}
    I_\nu(x,y,\lambda) =& \bigg[\sum_{i=1}^{2}A_i^{\rm dust}\frac{B_\nu(\lambda,T_i)}{(\lambda/9.7)^2} +
    \sum_{j=1}^{N_{\rm PAH}}A_j^{\rm PAH}D_j(\lambda) 
    + \sum_{k=1}^{N_{\rm line}}A_k^{\rm line}G_k(\lambda)\bigg] e^{-\tau(\lambda)}\\ 
    & + A^{\rm QSO}(\lambda)I_{\nu}^{\rm QSO}(\lambda){\rm PSF}(x,y,\lambda)
\end{split}
\end{align}

The first term represents thermal dust emission, where $B_\nu(\lambda, T_i)$ is the Planck function per unit frequency $\nu$ at temperature $T_i$, and we assume a dust emissivity $\propto 1/\lambda^2$. We fit two dust temperatures in the range (35, 300) K, where the upper limit is chosen to prevent fitting hotter dust temperatures that are presumably heated by the QSO.
% thus allowing the host galaxy and QSO to be consistently decomposed. 
This limit is also consistent with that used by [\citenum{2007ApJ...656..770S}] in the PAHFIT software to fit star-forming galaxies. The second term is the emission from PAH molecules, where $D_j(\lambda)$ is a Drude profile\cite{1986ApJ...307..286F}. The central wavelengths and FWHMs of the Drude profiles are taken from the average values obtained by [\citenum{2023MNRAS.519.3691D}], which are updated for JWST based on the original values from PAHFIT. We also allow a small variation of $\pm$0.05 $\mu$m for the centers and $\pm$10\% for the widths to account for the increased spectral resolution of MIRI compared to Spitzer. The third term is the gas emission lines, where $G_k(\lambda)$ is a Gaussian profile. Each line is allowed to be modeled by up to 2 Gaussian profiles (see the Supplementary Information). We correct the measured FWHMs using the FWHM of the line spread function (LSF) derived from Figure 3 of [\citenum{2021A&A...656A..57L}]. The LOS velocities are limited to $\pm$800 \kms, and the FWHMs are limited to $\leqslant$1500 \kms. We also tie the kinematics of lines with similar ionization potentials (IPs) to represent different gas phases, ending up with 4 groups: molecular H$_2$, cool (IP $<$ 24 eV), warm (24 $<$ IP $<$ 90 eV), and coronal (IP $>$ 90 eV). The wavelengths and IPs of the fit coronal lines, plus \neii\, \rev{and [S\,\textsc{iv}]}, are given in Extended Data Table \ref{tab:lines}. We assume a simple screen geometry such that these first three terms are extincted by a factor $e^{-\tau(\lambda)}$ where $\tau$ is the optical depth of silicate dust. We follow the methods of [\citenum{2021A&A...651A.117T}] to model 
$\tau(\lambda)$ by dividing the silicates into three species. We fix the relative abundances of these species based on a fit to the integrated spectrum from channels 1--4, since we find that allowing them to be free in each spaxel often leads to the extinction compensating for residual instrument artifacts and producing unphysical results. The final term represents the contamination from the QSO spectrum that is present due to the PSF (see the Supplementary Information). $I_\nu^{\rm QSO}$ is taken as the spectrum of the brightest spaxel divided by the PSF model of the same spaxel. The normalization, $A^{\rm QSO}(\lambda)$, is constant in each channel, but may be different between channels. This represents how far the QSO contribution is allowed to deviate from what is predicted purely from our PSF model, which we constrain within a factor of 2: (0.5, 2). 
% This allows the QSO model to more accurately fit the jumps in continuum level between channels.

\subsection{Modeling Procedure} 
\label{sec:method_spec_proc}

\hfill\break
Our minimization procedure is subdivided into 3 stages:
\begin{enumerate}
    \item Mask out the emission lines and fit the continuum. In this step, the PAHs are not fit with Drude profiles.  Rather, they are modeled using two of the templates from [\citenum{2007ApJ...656..770S}], as in [\citenum{2009ApJS..182..628V}]. This was done because the spectra have a relatively high optical depth and weak PAH features. Fitting everything simultaneously can thus lead to the spectral decomposition confusing how much of the flux around the 9.7 $\mu$m silicate absorption feature is due to the broad wings of the surrounding PAH features vs. the continuum, leading to unphysically large PAH fluxes. 
    % By fitting the PAHs with templates, this degeneracy is avoided.
    \item Subtract the thermal dust continuum and QSO contribution obtained from step 1, mask out the emission lines, and fit the PAHs to the residual spectrum using Drude profiles. 
    % This allows us to fit the PAHs more accurately and obtain individual fluxes for each feature without compromising the degeneracy mentioned in step 1.
    \item Mask out the emission lines, subtract the QSO template contribution, and fit a cubic spline to the residuals with a spacing of 7 pixels between knots. Subtract this cubic spline fit and use the residuals to fit the emission lines. By subtracting a cubic spline fit instead of the fit obtained from the previous steps, we are able to obtain accurate line fluxes even when the physically motivated continuum model under- or over-predicts the continuum locally around a line.
\end{enumerate}
\noindent
We use the Levenberg-Marquardt (LM) least squares minimization algorithm \cite{2009ASPC..411..251M} for each step in the fitting process. However, for the emission line fit in step 3, we use the simulated annealing (SA) global minimization algorithm as implemented by \texttt{Optim.jl} \cite{mogensen2018optim}, whose results are then used as a starting point for the LM minimization to refine the parameters.
% This helps in particular for fitting lines with more than 1 gaussian profile where different decompositions can produce similar overall profiles, and the noise in the spectrum can introduce local minima that do not produce the best overall models. We use the results of this fit as the starting point for a local minimization with LM to refine the results.

To generate our results, we implement this procedure to fit spaxel-by-spaxel on a data cube containing the combined channel 3 data, and similarly for the channel 2 and 4 data (cutting out channel 4C due to low sensitivity). Since the QSO PSF component outshines the host galaxy in many spaxels, this makes it difficult to measure the underlying host galaxy optical depth. As such, we \rev{have} developed a technique to measure this independently based on the flux ratios of the rotational H$_2$ lines. The H$_2$ 0-0 $S(3)$ line lands near the peak of the 9.7 $\mu$m silicate absorption feature, while the $S(4)$ line is unaffected, so we use the $S(4)/S(3)$ line ratio as a proxy for the optical depth. We do an initial fit to measure the observed H$_2$ line fluxes, then we fit excitation models to the upper-level column densities, masking out the $S(3)$ line, to obtain an ``intrinsic'' $S(4)/S(3)$ line ratio. Comparing this with the observed line ratio gives us the optical depth, which we then use to run a second iteration of our fitting procedure. For a more detailed description of this process, see Reefe et al. 2024, in prep. We also repeat all of these fits while not including the QSO template component (substituting it for an additional thermal dust component that may be up to 1500 K), to allow for a comparison with the results for the QSO-subtracted maps. For regions of interest, such as the northern extended cloud and the nucleus, we create an integrated 1D spectrum combining channels 2--4 (excluding 4C) and fit with 100 bootstrapping iterations. We also do this for the entire FOV of channel 2. See examples of 1D spectral fits in Extended Data Figures \ref{fig:1d_spectrum_all} and \ref{fig:1d_spectrum_ch3}. 2D maps of the fluxes, LOS velocities, and line widths for \nevi, \neii, and \nev\, are shown in Extended Data Figures \ref{fig:nevi_flux}--\ref{fig:nev_kinematics}.

\section{Surface Brightness Models}
\label{sec:res_surfacebrightness}

To create the radial surface brightness profile in Figure \ref{fig:radialflux}, we use a series of rectangular apertures of 3x1 spaxels going from the nucleus through the two cooling clouds, and averaging these together.  The outermost and innermost apertures become circular annuli to improve the statistics.  We then model this profile by considering the case where all of the \nevi\, emission can be attributed to the central QSO, assuming a constant gas density and a consistent emitting cloud geometry. In this scenario, we would expect an inverse-square law fall-off in emission line intensity.  As seen in Figure \ref{fig:radialflux}, there is a distinct bump that deviates away from the $r^{-2}$ power law at a distance of $\sim 5$--13 kpc from the nucleus, corresponding to the emission region beneath the X-ray bubble (at $\sim 20$ kpc). This bump could be explained by a local increase in the electron density ($n_e$), since the \nevi\, line emission, coming from collisional excitations, is proportional to $n_e n_{\rm Ne^{5+}}$.  However, the fact that we also see a bump in the \nevi/\nev\, ratio, which is insensitive to both the gas density $n_e$ and the metallicity, at the same location suggests that there is a local change in the ionization parameter ($U \equiv n_{\gamma,i}/n_{\rm H}$, the ratio of the ionizing photon density to the hydrogen density) and thus the dominant ionization mechanism.  To account for this, we add a second component to our model, representing the cloud as a simple boxcar function with varying position and width. Fitting this 2-component model, we find the cloud to be centered at $\sim$10 kpc and has a size along the radial direction of $\leqslant 5$ kpc---above this, the fit becomes significantly worse, as shown by the 10 kpc boxcar size in the Figure. Note that this model has also been convolved with the instrumental PSF (assuming it is a Gaussian). 

\section{Cooling Models}
\label{sec:res_cloudy}

We assume that the coronal emission comes from ICM gas cooling to low temperatures. In our simulations, starting from the ICM density and temperature discussed in the main text, we let a unit volume parcel of gas cool from the ambient temperature down to 10,000 K, and track its evolution assuming that it cools under either isobaric or isochoric conditions. We use version 23.00 of \textsc{Cloudy}\cite{2023RMxAA..59..327C} for our simulations, which has been updated to the more accurate $R$-matrix collision strength results of version 10 of the Chianti database\cite{2021ApJ...909...38D}. For each line, we accumulate the emissivities over temperature, denoted by $\Gamma$, as described in [\citenum{2015MNRAS.446.1234C}]. These are related to the observed line luminosities via $L_\mathrm{line} = \dot{M} \Gamma$, where $\dot{M}$ is the mass cooling rate.

A pure collisional model led to unphysically high mass cooling rates for all observed lines. We therefore also took into account photoionization by the radiation field of the central AGN, and performed a number of simulations at various distances from the AGN, in the range 2--20 kpc. To illustrate the gross influence of the AGN, it is worth noting that the gas fails to reach the target temperature, but instead comes into equilibrium with the radiation field at a higher temperature. In more detail, photoionization affects the ionization balance of the gas by driving long ionization tails that extend to much lower temperatures than in the purely collisional case.

How exactly the gas reacts to the AGN radiation depends on how it cools. In isobaric cooling, the effect of photoionization diminishes as the temperature drops due to compression. The gas that contributes to the emissivity of coronal lines, in particular \nev\, and \nevi, covers a much broader temperature range compared to purely collisional cooling, that is, the gas is overionized as it cools to nebular temperatures, and the integrated emissivities of all coronal lines of interest are boosted. By contrast, in isochoric (constant density) cooling the impact of photoionization is more significant, the gas comes to equilibrium with the AGN radiation at higher temperatures than the isobaric case, while its effect on integrated emissivities is more complicated, with some enhanced relative to the purely collisional case, and some reduced.

In our final runs, we used a combination of isobaric and isochoric cooling. For a description of these methods and our systematic uncertainty analysis, see  the Supplementary Information.

\section{Kinematic Models}
\label{sec:res_kinematics}

Here we expand on how the velocity models shown in Figure \ref{fig:velocity_profile} are created. Each model has a different inclination angle $i$ that is adjusted to best match the observed profile, such that $v_{\rm los} = v\sin i$. 

The solid red line simply gives the circular rotational velocity: 
\begin{equation}
    v = \sqrt{\frac{GM(r)}{r}}
\end{equation}
where $M(r)$ is obtained from a combination of strong lensing and X-ray analyses\cite{2019ApJ...885...63M}. The dashed red line gives a ballistic model where a test particle starts at $r_0 = 6.75$ kpc and $v_0\sin i = 270$ \kms\, and is then allowed to free fall. The blue line gives an ionized wind model driven by supernovae from [\citenum{2022ApJ...924...82F}], including the effects of gravity and cooling, and using a SFR of 800 \msunyr\, [\citenum{2012Natur.488..349M}], mass loading factor $\eta_M = 0.3$, and a half-opening angle $\pi/8$. The purple line gives an outflow model driven by radiation pressure from an AGN, using an analytic model from [\citenum{2023ApJ...943...98M}]:
\begin{equation}
    v(r) = \sqrt{\int_{r_1}^{r}\bigg[4885\frac{L_{44}\mathcal{M}}{r^2}-8.6 \times 10^{-3}\frac{M}{r^2}\bigg]{\rm d}r}
\end{equation}
with a launch radius of $r_1 = 1$ kpc, a bolometric AGN luminosity of $L_{44} = 3000$ ($\times 10^{44}$ \ergs)\cite{2012Natur.488..349M}, and force multiplier $\mathcal{M} = 500$, which accounts for the increased efficiency of radiative acceleration above that of pure Thomson scattering ($\mathcal{M} = 1$) from bound-bound, bound-free, and free-free opacities.

Finally, the bubble model, shown in green, is obtained by numerically solving a series of differential equations:
\begin{align}
    \dv{v}{t} &= \frac{\rho_{\rm ICM}V_{\rm b}}{M_{\rm b}}\frac{GM(r)}{r^2} - \frac{1}{2}\frac{\rho_{\rm ICM}}{M_{\rm b}}C_D A_{\rm b} v^2 \label{eq:bubble_velocity} \\
    \dv{V_{\rm b}}{t} &= -\frac{V_{\rm b}}{\gamma_{\rm ad} P_{\rm b}} \dv{P_{\rm b}}{t} \label{eq:bubble_volume}
    % \dv{P_{\rm b}}{t} &= \dv{P_{\rm b}}{r} v \\\label{eq:bubble_pressure}
    % \dv{A_{\rm b}}{t} &= 2\pi R_{\rm b} \dv{R_{\rm b}}{t} \label{eq:bubble_area} \\
    % \dv{R_{\rm b}}{t} &= \frac{1}{4\pi R_{\rm b}^2}\dv{V_{\rm b}}{t} \label{eq:bubble_radius} 
\end{align}
Here, $r$ is the distance from the nucleus to the center of the bubble and $v$ is the bubble velocity. The bubble is assumed to be spherical, with $V_{\rm b}$, $P_{\rm b}$, $A_{\rm b}$, $M_{\rm b}$, and $R_{\rm b}$ being the volume, pressure, cross-sectional area, mass, and radius of the bubble. $M(r)$ is the enclosed mass of the galaxy at radius $r$, $\rho_{\rm ICM} = \mu m_p n_e$ is the density of the ambient ICM, $C_D = 0.5$ is the drag coefficient, and $\gamma_{\rm ad} = 4/3$ is the adiabatic index of the bubble\cite{2010MNRAS.406.2023P}. Equation \eqref{eq:bubble_velocity} gives the velocity from the competing effects of the buoyancy force and drag force---the effects of gravity on the entrained mass are ignored. Equation \eqref{eq:bubble_volume} gives the rate of expansion of the bubble volume assuming adiabatic expansion. 
% Equations \eqref{eq:bubble_pressure}--\eqref{eq:bubble_radius} give the other properties of the bubble as it expands. 
For the pressure $P_{\rm b}(r)$ and density $\rho_{\rm ICM}(r)$ we use the 1D pressure and density profiles of the Phoenix cluster ICM obtained from \textit{Chandra} observations\cite{2015ApJ...811..111M}, from which we obtain $\dv*{P_{\rm b}}{r}$ using a finite difference approximation. 

The bubble starts at $r = 0.05$ kpc, whereas our lowest sample for the pressure and density profiles are at 3 kpc, which necessitates extrapolating down to the starting radius of the bubble. The thermodynamics at such small radii are not well constrained, so we take two extreme approaches: (1) assuming the pressure and density are both flat below 3 kpc, and (2) extrapolating the pressure and density below 3 kpc with a power law based on the first few data points. The starting parameters of both models are shown in \rev{Supplementary Information Table} \ref{tab:bubble}. The initial bubble radius is chosen such that it evolves to the currently observed size of the bubble ($\sim 8.5$ kpc)\cite{2015ApJ...805...35H} when it reaches the currently observed position of the bubble ($\sim 20$ kpc)\cite{2015ApJ...805...35H}, and the entrained mass is adjusted to best match the observed velocity profile.  Note that in addition to the uncertainty imparted by the extrapolation of the pressure and density profiles below 3 kpc, the model parameters listed in \rev{Supplementary Information Table} \ref{tab:bubble} have some large degeneracies with each other---in particular, the mass ($M_{\rm b}$) / density ($\rho_{\rm b,0}$) and inclination ($i$), which both affect the overall normalization of the velocity curve.  We stress that the purpose of these models is to illustrate the general agreement between the data and the bubble uplift scenario, but a rigorous model capturing the intricate physics expected in a bubble's wake would be far more complex. As such, we do not attempt to use these models to make any predictions or constraints on the actual amount of cooling coronal gas.

\clearpage
\renewcommand{\figurename}{Extended Data Figure}
\setcounter{figure}{0}
\renewcommand{\tablename}{Extended Data Table}
\setcounter{table}{0}

\begin{figure}
    \centering
    \includegraphics[width=0.5\textwidth]{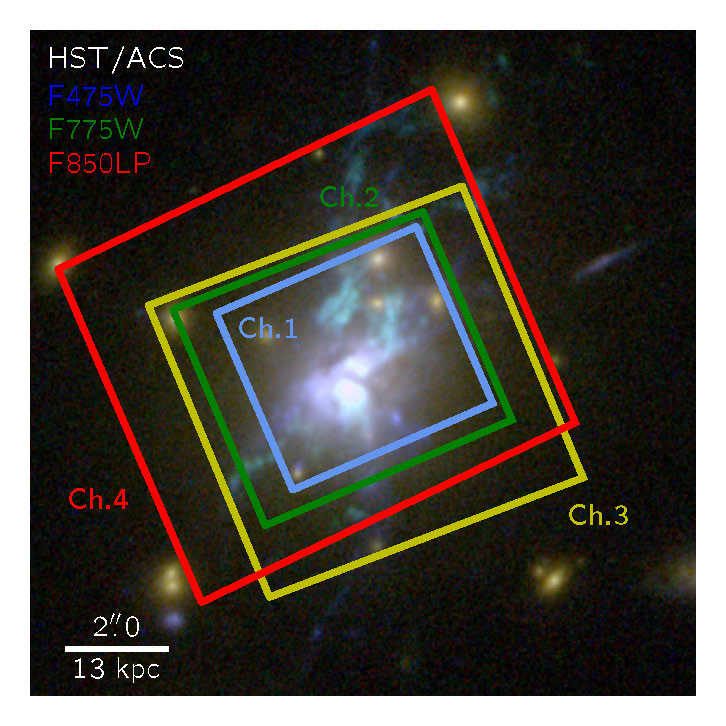}
    \caption{\textbf{The field of view of the MIRI/MRS instrument relative to the size of the central galaxy in the Phoenix Cluster.} The field of view of each of the four MIRI channels shown over a 3-color image of the core of the Phoenix cluster using data from HST/ACS in the F475W, F775W, and F850LP filters. The dust-obscured AGN lies at the center of the brightest cluster galaxy. The angular and physical scales are annotated in the bottom-left.}
    \label{fig:miri_fov}
\end{figure}

\begin{figure*}
    \centering
    \includegraphics[width=0.8\textwidth]{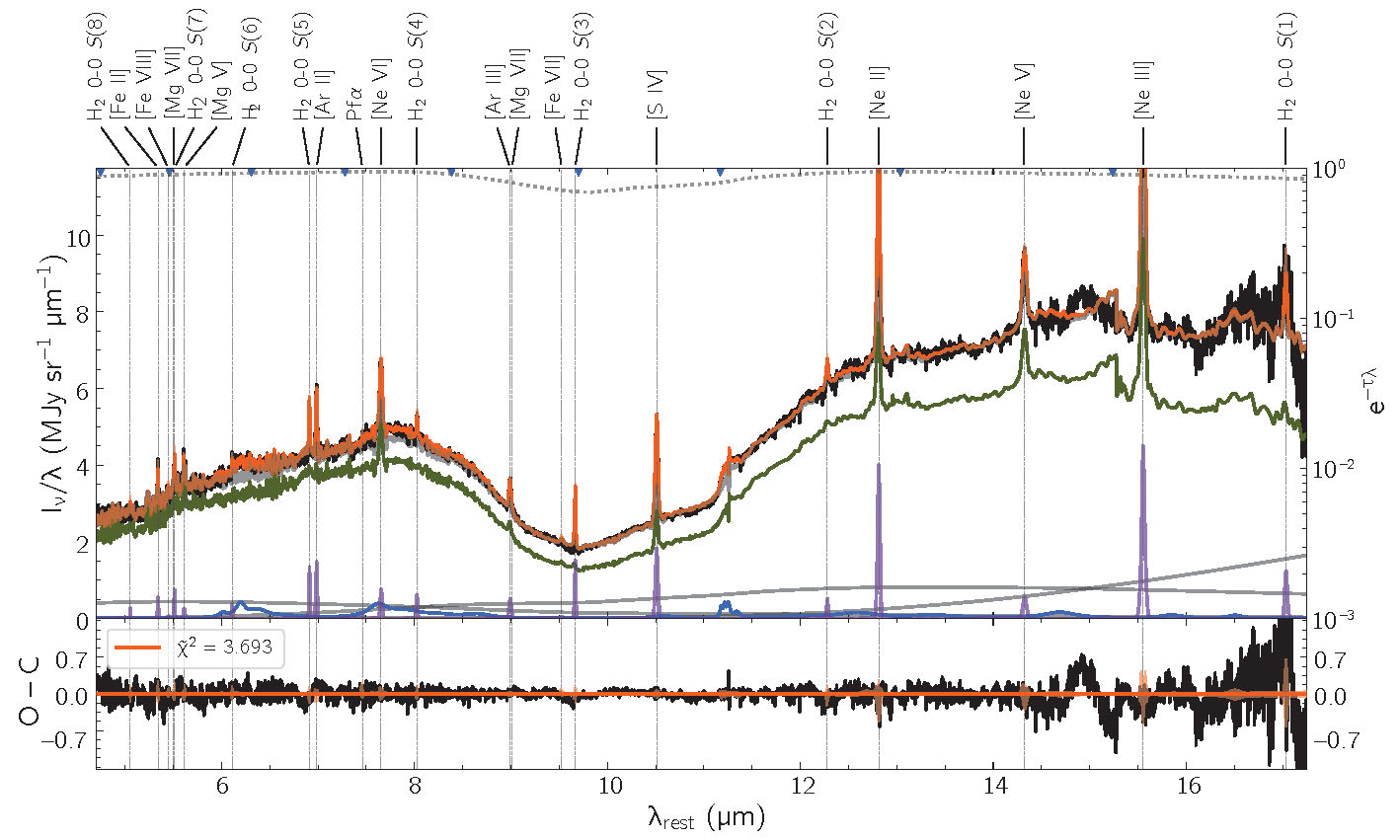}
    \caption{\textbf{The full mid-infrared spectrum of the central galaxy in the Phoenix Cluster.} A spectrum covering MIRI channels 2--4 (excluding the long wavelength end of 4C), and integrated over the full channel 2 FOV. The top panel shows the data in black and the model in orange, while the bottom panel shows the residuals. The model is also decomposed into components: the gray solid lines are the thermal dust continua, the green line is the QSO PSF model, the blue line is the PAH emission, the purple lines are the emission lines, and the dotted gray line shows the extinction profile. The sum of the thermal dust continua and QSO PSF is also shown by the thick gray line. The emission lines are labeled at the top of the plot, with vertical dashed lines showing their locations in the spectrum. The boundaries between channels and bands are also shown at the top of the plot with blue arrows. The translucent orange region shows the range of models produced in the different bootstrapping iterations}
    \label{fig:1d_spectrum_all}
\end{figure*}

\begin{figure*}
    \centering
    \includegraphics[width=0.8\textwidth]{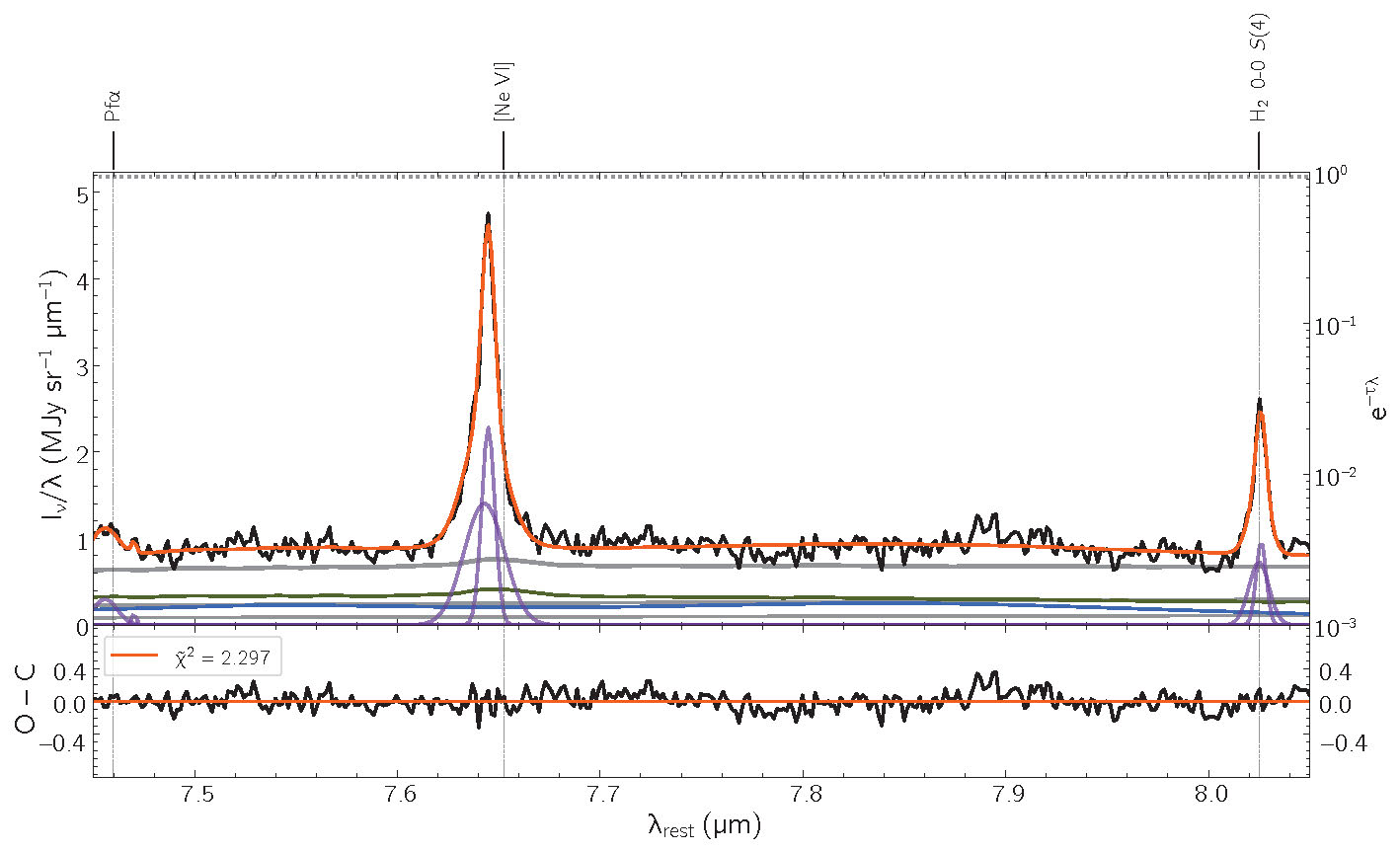}
    \caption{\textbf{A zoomed-in emission line spectrum of the central galaxy in the Phoenix Cluster.} A subset of the MIRI channel 3 spectrum zoomed in on the \nevi\, line. This spectrum includes only a single spaxel north of the nucleus. The formatting of everything is identical to Extended Data Figure \ref{fig:1d_spectrum_all}, except that there is no translucent orange region since this fit has not been bootstrapped. The two distinct kinematic components of the emission lines can be seen in purple.}
    \label{fig:1d_spectrum_ch3}
\end{figure*}

\begin{figure*}
    \centering
    \quad
    \includegraphics[width=\textwidth]{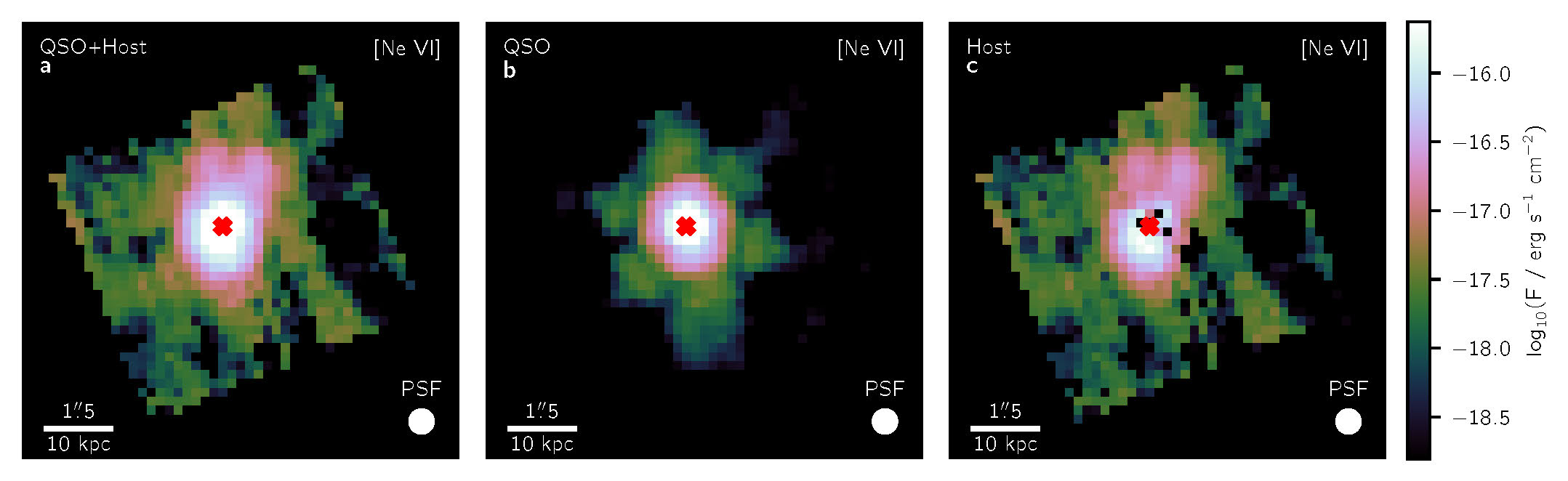}
    \includegraphics[width=\textwidth]{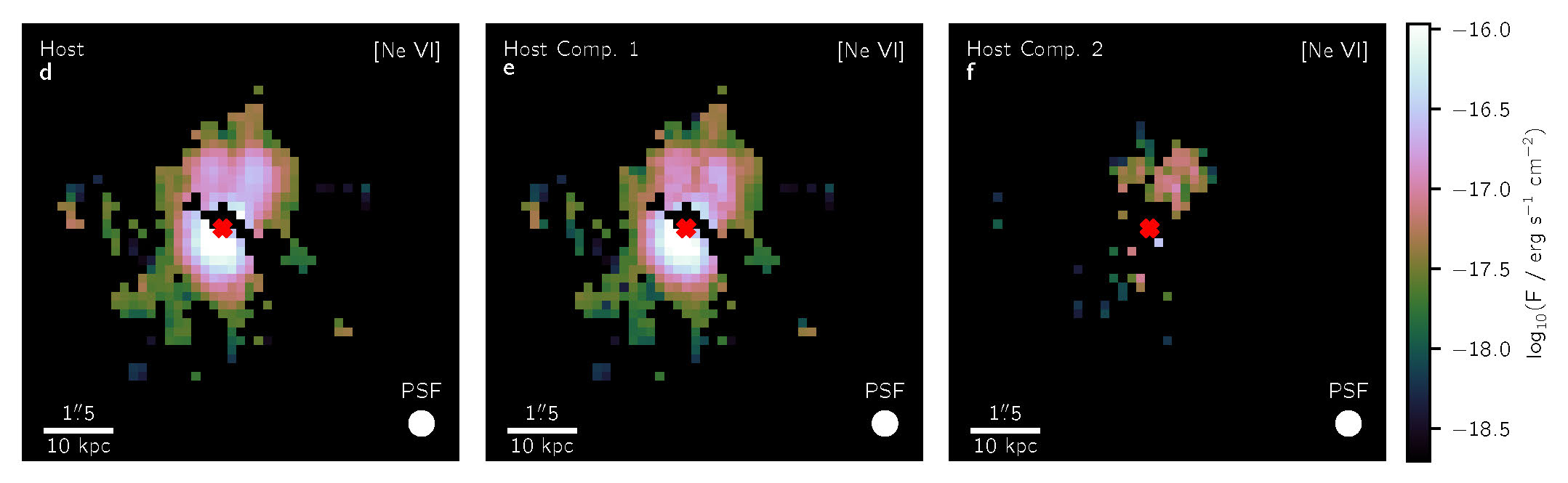}
    \caption{\textbf{A series of flux maps for \nevi~over the MIRI channel 3 FOV showing the decomposition into various spatial and kinematic components.} Panels (a-c) show the QSO and host galaxy components of the flux. (a) shows the total observed flux, (b) shows the flux from the QSO that has been dispersed according to the PSF model, and (c) shows the QSO-subtracted flux that is attributed to the host galaxy. Panels (d-f) show the further decomposition of the host galaxy flux into 2 distinct kinematic components, now with an $S/N$ cut such that only spaxels with a detection of $S/N \geqslant$ 3 are shown. (d) shows the combined flux from both kinematic components, making it identical to (c) except for the SNR cut. (e) and (f) show the fluxes from each individual kinematic component, sorted in order of decreasing flux. The color scales are shown on the right of each row and are the same for each panel in that row. The physical scale in kpc and angular scale in arcsec are annotated in the bottom left of each panel, and the FWHM of the PSF at the wavelength of \nevi~is shown in the bottom right of each panel.
    }
    \label{fig:nevi_flux}
\end{figure*}

\begin{figure*}
    \centering
    \quad
    \includegraphics[width=\textwidth]{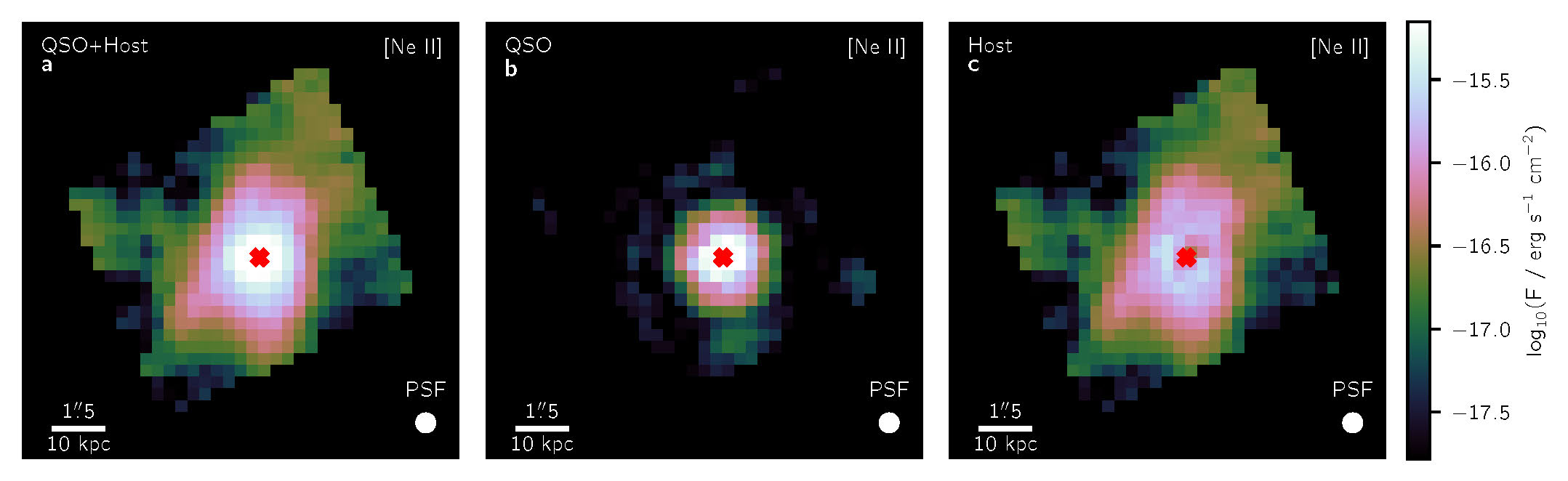}
    \includegraphics[width=\textwidth]{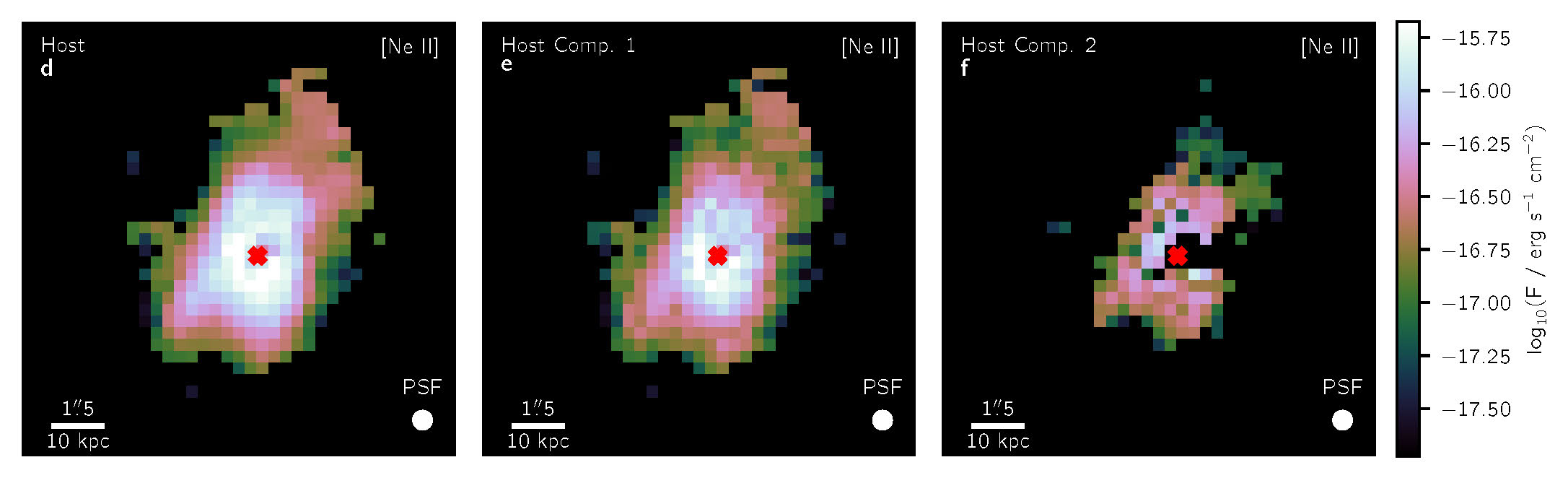}
    \caption{\textbf{A series of flux maps for \neii~over the MIRI channel 4 FOV showing the decomposition into various spatial and kinematic components.} Same as Extended Data Figure \ref{fig:nevi_flux}, but for \neiiwave.}
    \label{fig:neii_flux}
\end{figure*}

\begin{figure*}
    \centering
    \quad
    \includegraphics[width=\textwidth]{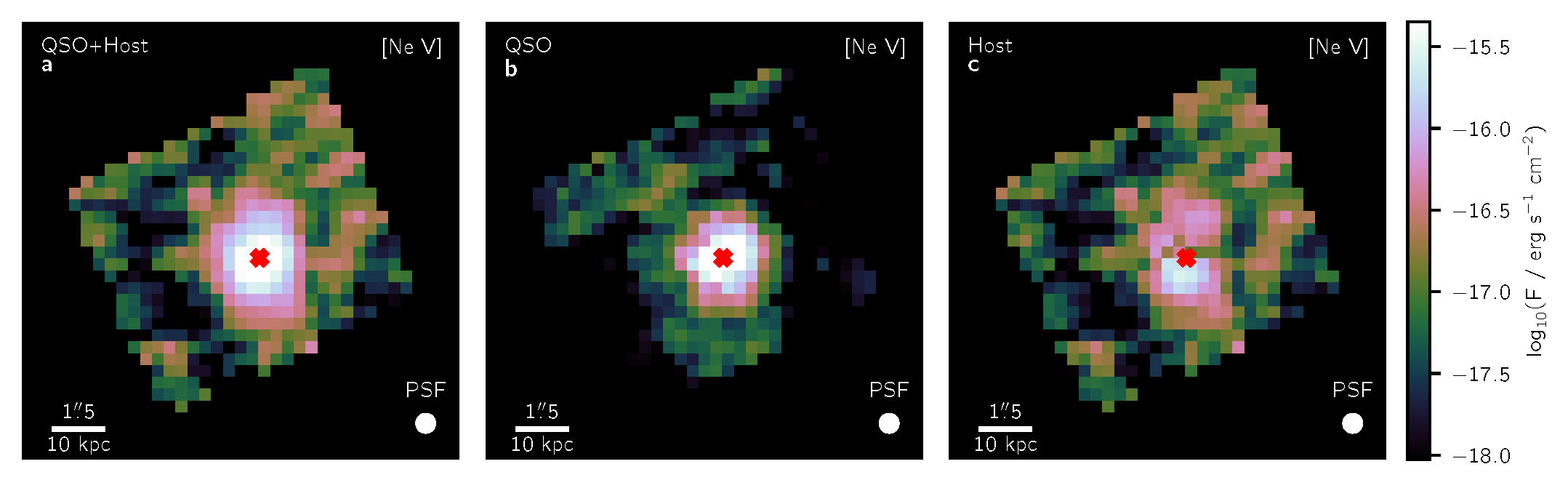}
    \includegraphics[width=\textwidth]{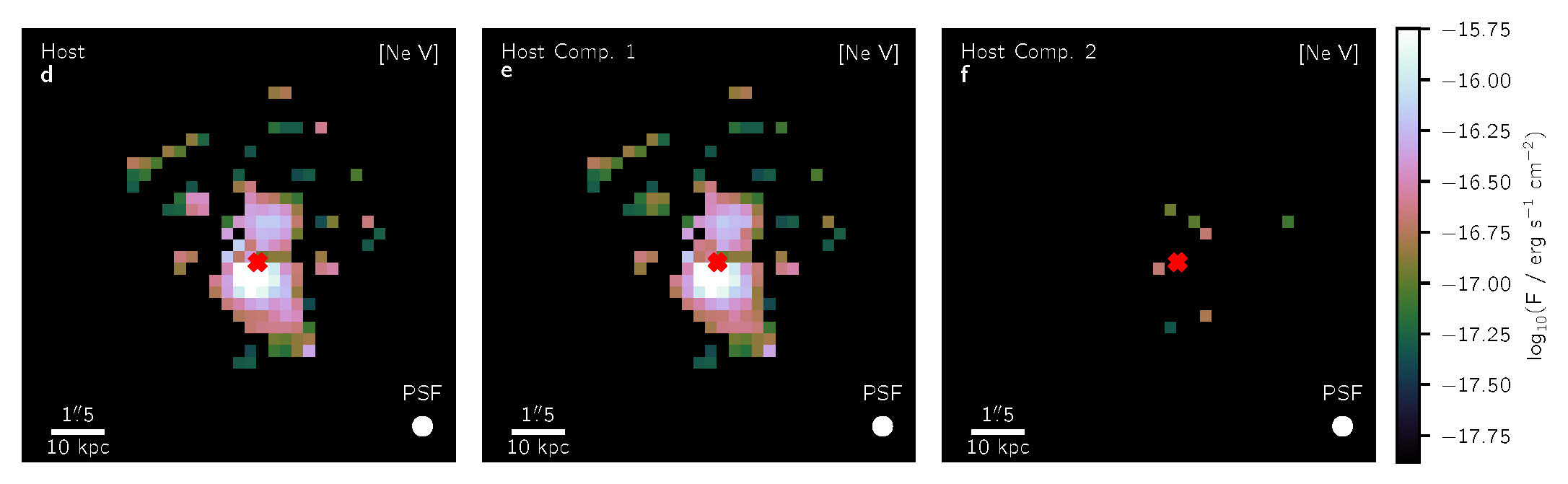}
    \caption{\textbf{A series of maps for \nev~over the MIRI channel 4 FOV showing the decomposition into various spatial and kinematic components.} Same as Extended Data Figure \ref{fig:nevi_flux}, but for \nevwave.}
    \label{fig:nev_flux}
\end{figure*}

\begin{figure*}
    \centering
    \quad
    \includegraphics[width=\textwidth]{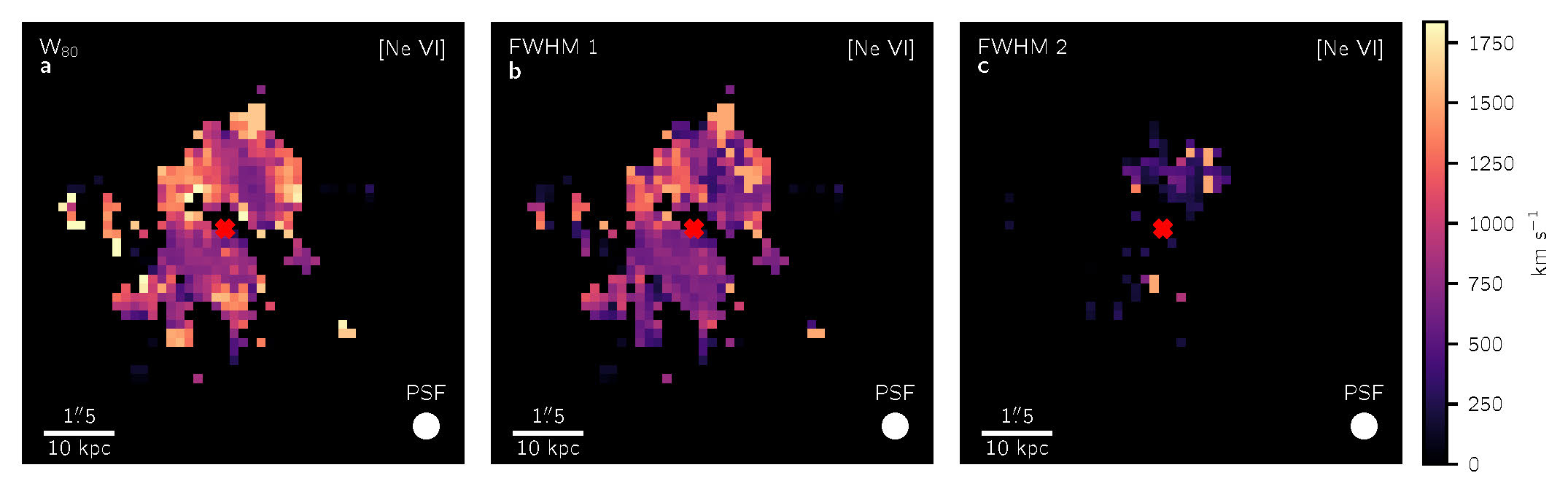}
    \includegraphics[width=\textwidth]{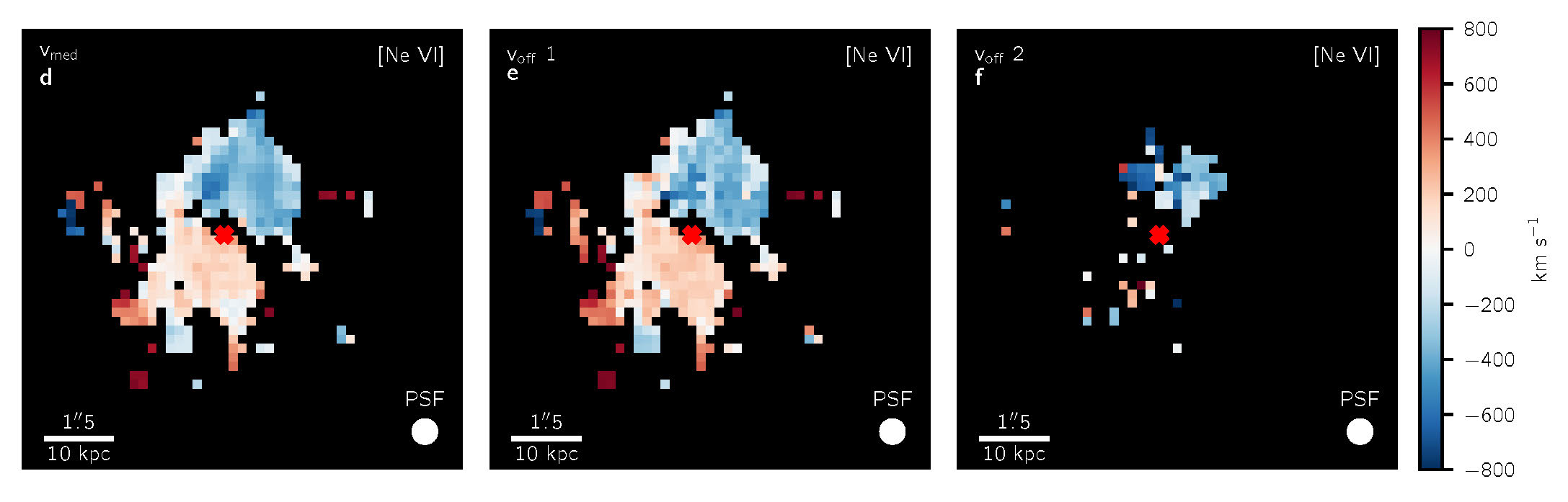}
    \caption{\textbf{A series of kinematic maps for \nevi~over the MIRI channel 3 FOV showing the decomposition into various kinematic components.} Panels (a-c) show the line velocity widths. (a) shows the $W_{80}$ of the full line profile, while (b) and (c) show the FWHMs of the individual kinematic components. Panels (d-f) show the LOS velocities. (d) shows the median LOS velocity of the full line profile, while (e) and (f) show the peak velocities of the individual kinematic components. All maps have an $S/N \geqslant 3$ cut. The color scales are shown on the right of each row and are the same for each panel in that row. The physical scale in kpc and angular scale in arcsec are annotated in the bottom left of each panel, and the FWHM of the PSF at the wavelength of \nevi~is shown in the bottom right of each panel.}
    \label{fig:nevi_kinematics}
\end{figure*}

\begin{figure*}
    \centering
    \quad
    \includegraphics[width=\textwidth]{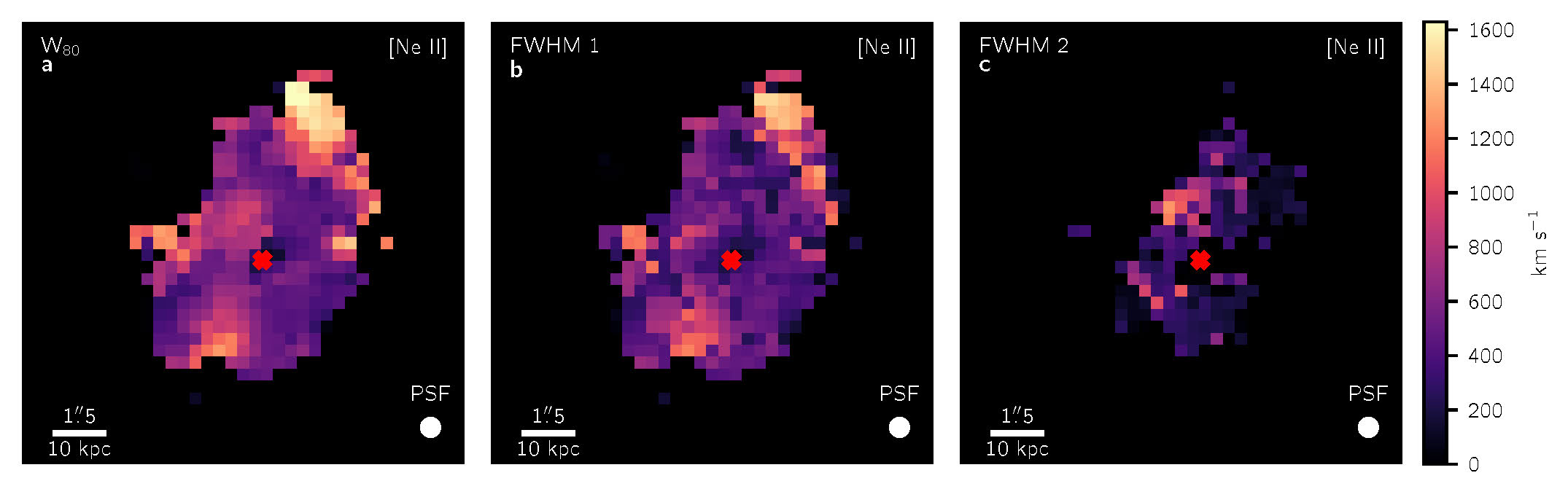}
    \includegraphics[width=\textwidth]{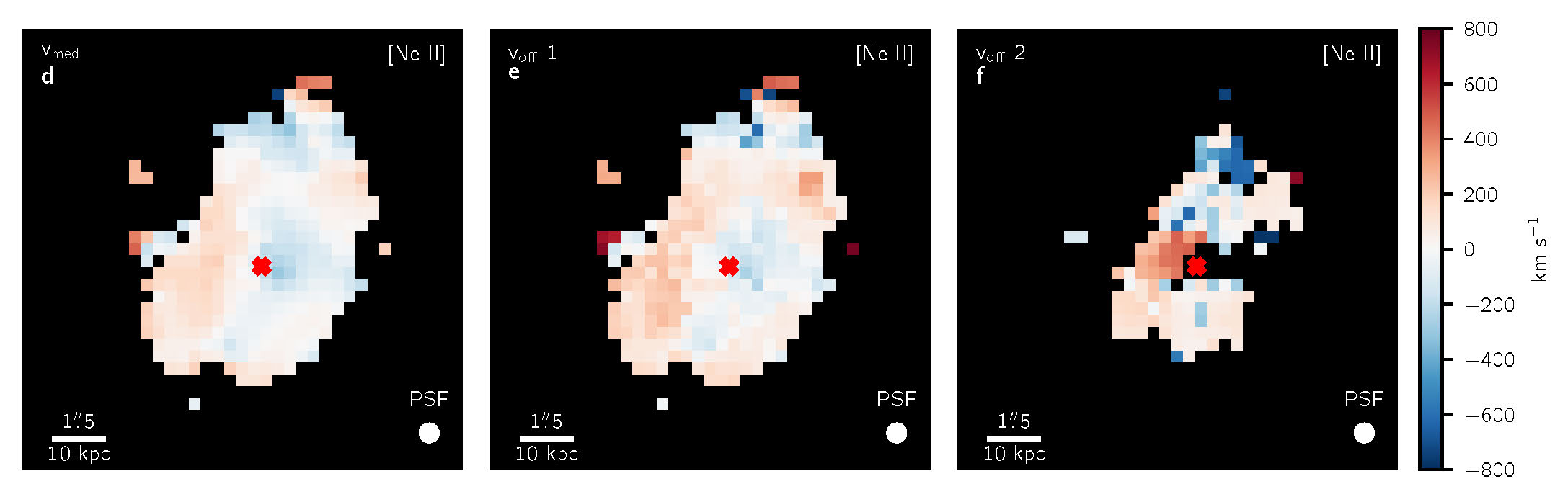}
    \caption{\textbf{A series of kinematic maps for \neii~over the MIRI channel 4 FOV showing the decomposition into various kinematic components.} Same as Extended Data Figure \ref{fig:nevi_kinematics}, but for \neiiwave.}
    \label{fig:neii_kinematics}
\end{figure*}

\begin{figure*}
    \centering
    \quad
    \includegraphics[width=\textwidth]{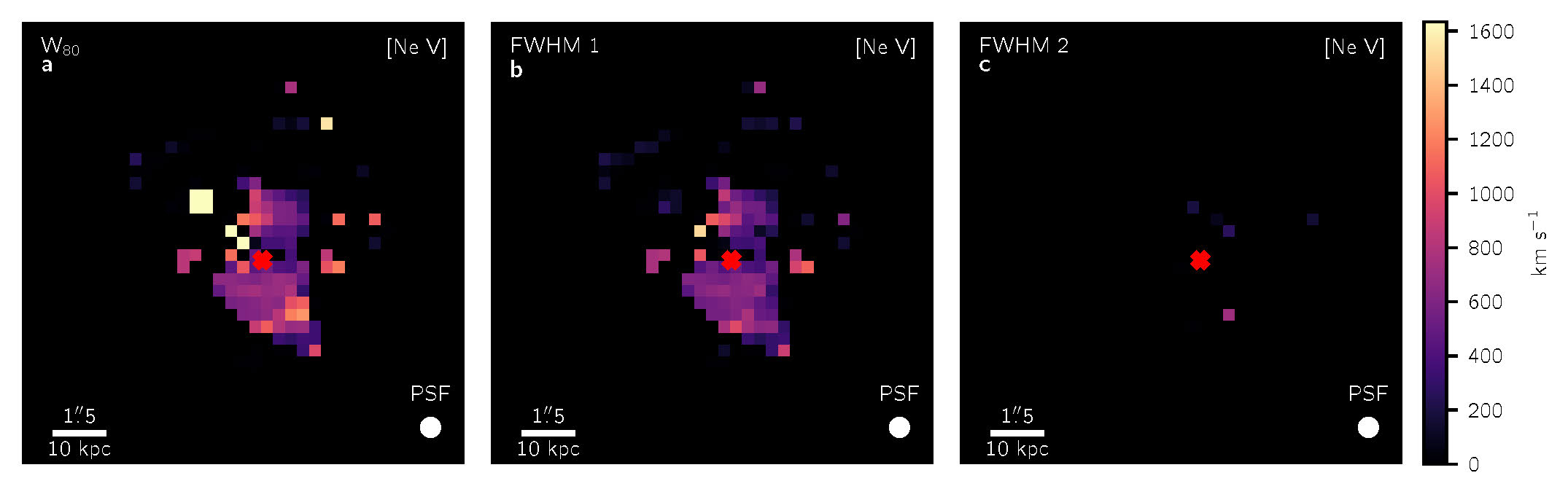}
    \includegraphics[width=\textwidth]{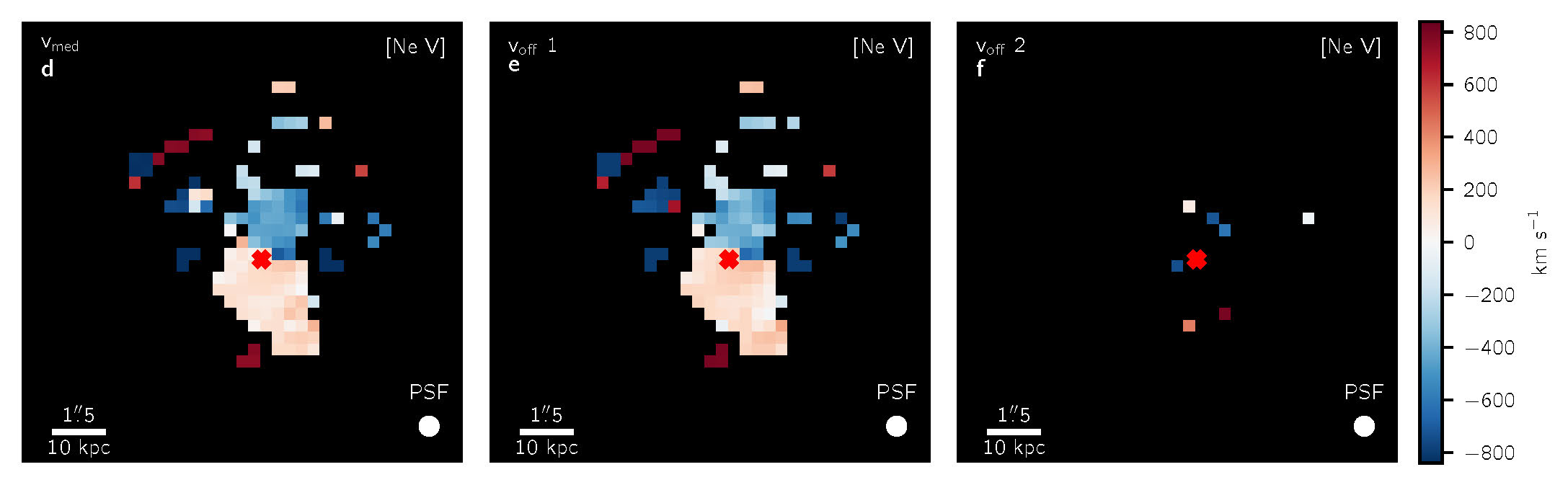}
    \caption{\textbf{A series of kinematic maps for \nev~over the MIRI channel 4 FOV showing the decomposition into various kinematic components.} Same as Extended Data Figure \ref{fig:nevi_kinematics}, but for \nevwave.}
    \label{fig:nev_kinematics}
\end{figure*}

\begin{table*}
    \caption{\textbf{Emission line fluxes in various regions}}
    \label{tab:lines}
    \vspace{0.1cm}
    \includegraphics[width=\textwidth]{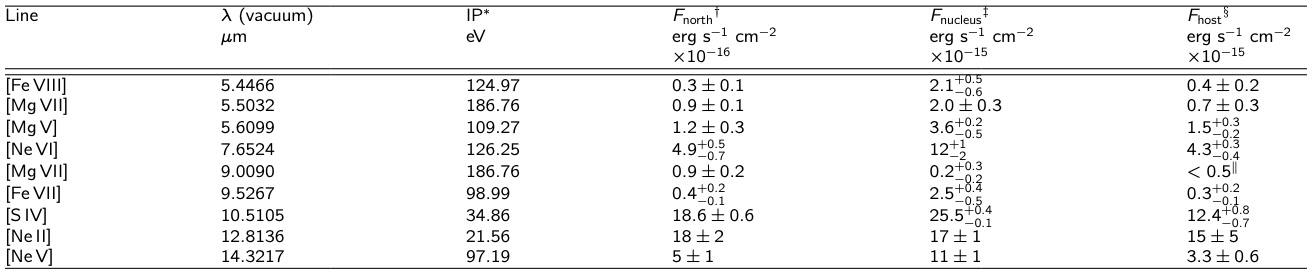}
    % \begin{tabular}{llllll}
    % \hline
    % {Line} & {$\lambda$ (vacuum)} & {IP$^\dagger$} & {$F_{\rm north}$$^1$} & {$F_{\rm nucleus}$$^2$} & {$F_{\rm host}$$^3$} \\
    % & $\mu$m & eV & \ergscm & \ergscm & \ergscm \\
    % & & & $\times 10^{-16}$ & $\times 10^{-15}$ & $\times 10^{-15}$ \\
    % \hline
    % \hline
    % % O VI & 0.1037 & 113.90 & - & - & - \\
    % \lbrack Fe\,\textsc{viii}\rbrack & 5.4466 & 124.97 & $0.3 \pm 0.1$ & $2.1^{+0.5}_{-0.6}$ & $0.4 \pm 0.2$ \\
    % \lbrack Mg\,\textsc{vii}\rbrack & 5.5032 & 186.76 & $0.9 \pm 0.1$ & $2.0 \pm 0.3$ & $0.7 \pm 0.3$ \\
    % \lbrack Mg\,\textsc{v}\rbrack & 5.6099 & 109.27 & $1.2 \pm 0.3$ & $3.6^{+0.2}_{-0.5}$ & $1.5^{+0.3}_{-0.2}$ \\
    % \lbrack Ne\,\textsc{vi}\rbrack & 7.6524 & 126.25 & $4.9^{+0.5}_{-0.7}$ & $12^{+1}_{-2}$ & $4.3^{+0.3}_{-0.4}$ \\
    % \lbrack Mg\,\textsc{vii}\rbrack & 9.0090 & 186.76 & $0.9 \pm 0.2$ & $0.2^{+0.3}_{-0.2}$ & $<0.5^*$ \\
    % \lbrack Fe\,\textsc{vii}\rbrack & 9.5267 & 98.99 & $0.4^{+0.2}_{-0.1}$ & $2.5^{+0.4}_{-0.5}$ & $0.3^{+0.2}_{-0.1}$ \\
    % \lbrack S\,\textsc{iv}\rbrack & 10.5105 & 34.86 & $18.6 \pm 0.6$ & $25.5^{+0.4}_{-0.1}$ & $12.4^{+0.8}_{-0.7}$ \\
    % \lbrack Ne\,\textsc{ii}\rbrack & 12.8136 & 21.56 & $18 \pm 2$ & $17 \pm 1$ & $15 \pm 5$ \\
    % \lbrack Ne\,\textsc{v}\rbrack & 14.3217 & 97.19 & $5 \pm 1$ & $11 \pm 1$ & $3.3 \pm 0.6$ \\
    % \hline
    % \end{tabular}
    \begin{tablenotes}
        \item[1] $^*$The ionization energy required to create ions of the given species, obtained from \href{https://physics.nist.gov/PhysRefData/ASD/ionEnergy.html}{NIST}.
        \item[2] $^\dag$The integrated flux within an elliptical aperture covering the region of extended emission to the north of the nucleus, with the QSO contribution subtracted.
        \item[3] $^\ddag$The integrated flux within a circular aperture with a $1''$ radius centered on the nucleus, including the QSO contribution.
        \item[4] $^\S$The integrated flux throughout the full channel 2 FOV, with the QSO contribution subtracted.
        \item[5] $^\parallel$$3\sigma$ upper limits are derived by integrating a gaussian with an amplitude of 3 times the continuum RMS at the location of the line, with a width of 1000 \kms.
    \end{tablenotes}
\end{table*}

\clearpage

\noindent{\large{\bf Supplementary Information}}
\setcounter{section}{0}

\renewcommand{\figurename}{Supplementary Information Figure}
\setcounter{figure}{0}
\renewcommand{\tablename}{Supplementary Information Table}
\setcounter{table}{0}

\section{Additional Methods for Data Cleaning and Spectral Modeling}

\subsection{Data Corrections}
\label{sec:obs_clean}
\hfill\break
After the main data reduction pipeline, we perform a few additional corrections to the data.  We remove residual stripe artifacts which exist along single rows of the IFU-aligned cubes by again following the procedure of [\citenum{2023Natur.618..708S}]: estimating the striping level in each detector row after masking out the bright central source in a circular aperture. This method relies on the fact that the spaxels outside the aperture do not contain significant source emission and are dominated by background. In our case, the galaxy emission exists essentially across the entire channel 1 FOV. However, channels 2--4 all contain strips to the sides of the galaxy that do not contain significant emission from the galaxy itself (see Extended Data Figure \ref{fig:miri_fov}) which can be used. Nevertheless, the extended emission from both the galaxy itself and the PSF of the central QSO contaminate a large portion of the FOV, leaving us with a small number of spaxels to estimate the background from in each row.  As such, we make some modifications to the original process to be careful not to accidentally remove real source emission: 1) we do not allow a slope in the striping level in each row, instead forcing each row to have a single value; 2) we subtract the median value from the stripe ``backgrounds'' at each wavelength slice before subtracting them from the data, such that on average we are not adding or subtracting any overall flux. These steps help to ensure that we are not inadvertently removing any real source emission.

Additionally, the MIRI team has noted a wavelength and time-dependent sensitivity drop, causing fluxes extracted from simulated WISE bandpasses to be underpredicted compared to WISE data (\href{https://jwst-docs.stsci.edu/jwst-calibration-pipeline-caveats/jwst-miri-mrs-pipeline-caveats}{https://jwst-docs.stsci.edu/jwst-calibration-pipeline-caveats/jwst-miri-mrs-pipeline-caveats}). This was corrected in recent versions of the JWST pipeline, but comparing simulated WISE fluxes from our final data cubes to WISE measurements still shows a discrepancy. As such, we saved the corrections the JWST was applying to each channel, fit a cubic polynomial, and exponentially scaled the corrections until the simulated JWST--WISE measurements were within 1$\sigma$ of the true WISE values. 
% For our data, this lead to an exponential scaling factor of 0.75.

The MIRI team has also noted that the errors produced by the pipeline are underestimated, sometimes up to a factor of 50. We have replaced these errors with our own calculations based on the data. In each spaxel, we estimate the point-by-point scatter in the data by finding the residuals between the data and a cubic spline fit, with knots spaced apart by 7 pixels and the emission lines masked out, and taking the standard deviation in a rolling window with a width of 60 pixels. 
% We find that this provides a more realistic estimation of the error than the pipeline values.

\subsection{Combining Data from Multiple Channels}
\label{sec:obs_combine}
\hfill\break
We have developed our own procedure to combine data cubes from different bands and channels into a single cube. First, the WCS parameters of each cube are adjusted to match each other based on the centroid positions of the edge regions where the channels overlap (these adjustments are often small, $\leqslant$0.2 pixels). Each cube is then projected onto the same grid, using the smallest FOV and smallest pixel scale from the input cubes. Finally, the spectra in the overlapping regions are resampled to a median resolution while conserving flux. Notably, combining data from multiple bands and channels often leads to discontinuous jumps in the continuum flux level at the channel boundaries. Our data contains a bright point source due to the QSO at the center of the galaxy, so we attribute the main source of these jumps to be differences in the size and shape of the PSF between the channels.

\subsection{PSF Template} 
\label{sec:method_spec_psf}
\hfill\break
We construct the PSF template by using observations of the bright star 16 Cygni B from program ID 1538. 
% We follow a similar data reduction procedure as detailed in ${\S}$\ref{sec:obs_reduce}. 
The pipeline version used for this reduction is 1.11.4 with CRDS context \texttt{jwst\_1118.pmap}. Many of the unique settings and corrections we used for the reduction of our Phoenix A data are no longer necessary since we are now in a much higher surface brightness regime. As such, we follow all of the default pipeline options. We then perform a residual background subtraction by subtracting the average flux within an annulus from 5--10x the PSF FWHM.

To create our PSF model from these observations, we shift the centroid of the 16 Cygni B observations to match the centroid position (in detector coordinates) of the Phoenix A observations and perform a cubic spline fit in each spaxel with a width of 100 pixels between each knot in the wavelength dimension, since the PSF ought to vary gradually in the wavelength axis. We also normalize the PSF such that it integrates to 1 at each wavelength slice, which removes the spectral shape of the star. This star is heavily affected by a spectral leak artifact at $\sim$12.2 $\mu$m (\href{https://jwst-docs.stsci.edu/known-issues-with-jwst-data/miri-known-issues/miri-mrs-known-issues}{https://jwst-docs.stsci.edu/known-issues-with-jwst-data/miri-known-issues/miri-mrs-known-issues}), whereas the Phoenix A data is relatively unaffected,
% (the effect leaks flux from channel 1, so red sources are much less affected than blue sources)
so we mask out and interpolate the data between 11.93--12.39 $\mu$m with a cubic polynomial.

\subsection{Line Component Testing}
\label{sec:method_spec_ftest}
\hfill\break
As mentioned in ${\S}$\ref{sec:method_spec_decomp}, we allow emission lines to be fit with either 1 or 2 Gaussian components. To determine whether a second component is statistically significant, we perform an $F$-test, requiring a threshold of 0.3\% (or 3$\sigma$) for the 2-component model to be favored over the 1-component model. Since we treat the line kinematics as tied into groups, for each group we choose the highest SNR line to undergo the $F$-test, to be used as a proxy for the kinematics of the entire group. If this test favors 2 components, then we fit all lines in the same group with 2-components, even if some of the other lines on their own would not pass the 3$\sigma$ threshold on the $F$-test. 
% The line chosen for the $F$-test also depends on the channel being fit. For channel 3, the chosen lines for each group are H$_2$ 0-0 $S(3)$, Pf$\alpha$, [S IV] $\lambda$10.511$\mu$m, and \neviwave. For channel 4, they are H$_2$ 0-0 $S(2)$, \neiiwave, and \nevwave~for the molecular, cool, and coronal groups (there are no warm group lines between 24--97 eV in channel 4A-B).

\section{\rev{Morphological Comparisons}} \label{sec:morph_compare}
% maybe move this(?) idk where to put it b/c we're at the word limit in the main text and the methods section

\rev{
It is interesting to note the morphological similarities and differences between the $10^{5.5}$ K coronal gas (Extended Data Figure \ref{fig:nevi_flux}), the $10^4$ K warm gas seen with Gemini\cite{2014ApJ...784...18M} (and the \neii\, line, Extended Data Figure \ref{fig:neii_flux}), and the 10 K molecular gas seen with ALMA\cite{2017ApJ...836..130R}. The hotter gas phases seem to be more centrally concentrated than the molecular gas, likely due to the influence of the AGN in the core causing them to be photoionized, but they generally retain the same shape, with unresolved filaments extending to the north, south, and southeast. These filaments, most prominently seen in the molecular gas, are draped around the edges of the X-ray cavities to the north and south, hinting that they are related to either the uplift of low entropy gas from the core or in situ formation of cooling gas in the turbulent wakes of the buoyantly rising bubbles. The peak of the \nevi\, emission (denoted by the red ``X'') appears to be offset from the peak of the optical [O \textsc{ii}] emission in Figure \ref{fig:overlay_city}, but we believe this is a result of the large dust extinction in the nucleus, as the peak of the \neii\, emission, which traces similar temperature gas as the [O \textsc{ii}] emission, is consistent with the peak of the \nevi\, emission. There is also a strong spatial correspondence between the axis of the northern radio jet emitted by the central AGN and a depression in the northern region of the extended high-ionization line emission, as shown by data from the Very Large Array (VLA), see Supplementary Information Figure \ref{fig:radiojet}.  This suggests that the jet plays a significant role in shaping the morphology of this gas, punching through it and pushing it away from the jet axis.  This depression is seen in the lower-ionization gas as well. The cospatiality and similar morphologies between phases indicates that cooling has likely been happening all throughout the inner 10s of kpc of the cluster within the recent past. 
}

\begin{figure*}
    \centering
    \includegraphics[width=0.5\textwidth]{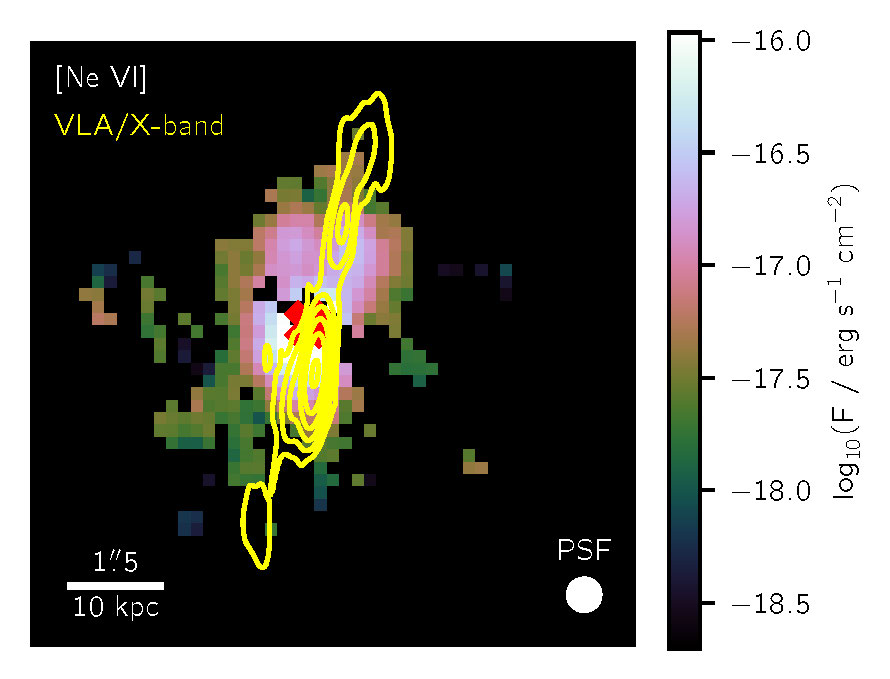}
    \caption{\textbf{A map of the \nevi-emitting coronal gas in the central galaxy of the Phoenix Cluster overlaid with the locations of the radio jets emitted by the AGN.} The QSO-subtracted \nevi~flux, with a cut of $S/N \geqslant 3$, is shown. VLA $X$-band data\cite{2021A&A...646A..38T} showing the trajectory of the radio jets are overlaid in yellow. The contours are at 2--9$\sigma$ above the background level measured in log space.}
    \label{fig:radiojet}
\end{figure*}

\rev{
The kinematics are also similar between the coronal gas (Figure \ref{fig:velocity_profile}, Extended Data Figure \ref{fig:nevi_kinematics}), the warm ionized gas (Extended Data Figure \ref{fig:neii_kinematics}), and the cold molecular gas\cite{2017ApJ...836..130R}, but with some interesting differences.  The cooler gas phases have systematically lower line widths and velocity gradients appearing in the East-West and North-South directions along the filaments, whereas the coronal gas has higher line widths and a primarily North-South velocity gradient.  Overall, this suggests that the molecular and warm phases may be more quiescent than the hotter gas phases in the filaments, which would be consistent with the expectations of the Chaotic Cold Accretion (CCA) model (see Ref. \citenum{2020NatAs...4...10G} for a review). However, the scales on which these three phases are probed via Gemini, ALMA, and \textit{JWST} are very different, so we defer a proper treatment to a follow-up paper in which we will probe all three phases (H$_2$, [Ne\,\textsc{ii}], [Ne\,\textsc{vi}]) with a single instrument.
% In this picture, there is a top-down cascade of condensation which ends in the molecular gas condensing out of the warm gas phase into individual clouds and filaments. These clouds are prevented from gaining mass through turbulent dissipation and they lose angular momentum through collisions with other clouds, causing them to rain onto the core and feed the central SMBH\cite{2013AN....334..394G, 2017MNRAS.466..677G, 2018ApJ...854..167G}. The higher line widths in the hotter phases may also be consistent with this picture, being influenced by turbulence imparted by the bubbles, but they are likely artificially broadened by capturing multiple clouds in a single spaxel (1 spaxel has a physical size of 1.35 kpc in channel 3), allowing for the possibility of low entropy gas being uplifted as well.
}

\section{Additional Methods for Cooling Models and Analysis}

\subsection{\rev{Composite Isobaric and Isochoric Cooling}} \label{sec:mixedcooling}
\hfill\break
Our simulations of cooling gas in the presence of AGN feedback, as presented in the main text, contain a fraction of isochoric and isobaric cooling\cite{2015MNRAS.451L..60G}.  To achieve this, we first run two simulations where the gas cools purely isobarically and purely isochorically, and obtained values for $\dot{M}$ and $\Gamma$ in each case.  Then, to find the right ratio of isochoric-to-isobaric cooling, we perform an optimization procedure on the line ratio diagram and find the ratio that minimizes the $\chi^2$ from the 1:1 line. We exclude the \neii\, and O\,\textsc{vi} lines from this calculation. We take the luminosity ratio of two emission lines in the \rev{composite} scenario to be
\begin{equation}
    \frac{L_1}{L_2} = \frac{\dot{M}_1}{\dot{M}_2}\frac{x\Gamma_{\rm c,1}+(1-x)\Gamma_{\rm b,1}}{x\Gamma_{\rm c,2}+(1-x)\Gamma_{\rm b,2}}
\end{equation}
where the subscripts ``b'' and ``c'' refer to isobaric and isochoric cooling, respectively, and $x$ is the fraction of isochoric cooling. We assume the $\dot{M}$ values of both lines are similar enough that their ratio can be approximated as 1. After finding the optimal value for $x$ (0.13), we use the new \rev{composite} $\Gamma$ values to convert the observed line luminosities into cooling rates:
\begin{equation}
    \dot{M}_{\rm obs} = \frac{L_{\rm obs}}{x\Gamma_{\rm c} + (1-x)\Gamma_{\rm b}}
\end{equation}
The resultant cooling rates are shown in \rev{Supplementary Information Table} \ref{tab:coolingrates}. For details on how we calculate systematic uncertainties, see Supplementary Information.

\subsection{\rev{Systematic Uncertainties}}
\label{sec:res_systematics}
\hfill\break
Although the statistical uncertainties quoted for the cooling rates suggest that these values are tightly constrained, in reality we are dominated by systematic uncertainties.  Two of the biggest factors in these systematic uncertainties are the bolometric luminosity of the quasar $L_{\rm bol}$ and the hot phase neon abundance $Z_{\rm Ne}$.  The former is difficult to constrain due to the highly obscured nature of the quasar and the potentially different obscuration along the line of sight of the cooling gas compared to ours.  The latter is difficult to constrain due to the limited spectral and spatial resolution of high-resolution X-ray spectroscopy, with the values we used being averaged over the inner 50$''$ (or 300 kpc) of the cluster.  We attempt to quantify these uncertainties by running a series of reference simulations in which everything is identical to our main simulations except for the value of one parameter ($L_{\rm bol}$ or $Z_{\rm Ne}$).  

Each parameter is adjusted to half and twice its nominal value. The cooling rates of each of these simulations are tabulated in \rev{Supplementary Information Table} \ref{tab:coolingrates}. This gives us a set of 3 data points which we can use to locally estimate the scaling between the cooling rates of each line and the adjusted parameters, i.e. $\dot{M}_i \propto L_{\rm bol}^a Z_{\rm Ne}^b$, where the exponents $a$ and $b$ are found by comparing the change in cooling rate to the change in each parameter.  For example, we find for \nevi\, a scaling of $\dot{M}_{\rm \nevi} \propto L_{\rm bol}^{-0.46}Z_{\rm Ne}^{-0.62}$. We then use this scaling to estimate the full bounds of systematic uncertainty in the case where \textbf{both} $L_{\rm bol}$ and $Z_{\rm Ne}$ are half or twice as large as their nominal values.  This gives us the range in systematic uncertainties quoted in the main text.

\begin{table*}
    \caption{\textbf{Emission line cooling rates and uncertainties}}
    \label{tab:coolingrates}
    \vspace{0.1cm}
    \includegraphics[width=\textwidth]{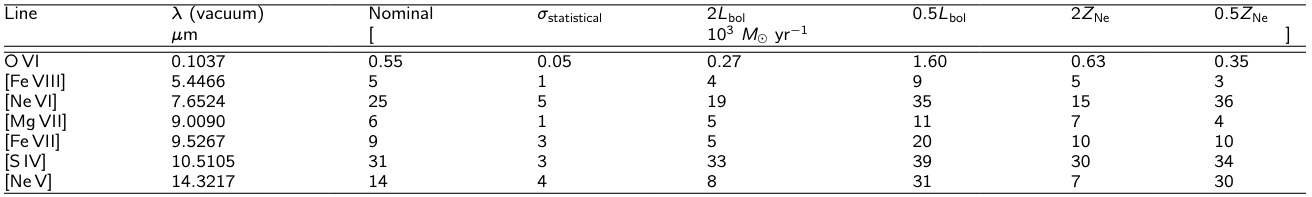}
    % \begin{tabular}{llllllll}
    % \hline
    % {Line} & {$\lambda$ (vacuum)} & Nominal & $\sigma_{\rm statistical}$ & $2L_{\rm bol}$ & $0.5L_{\rm bol}$ & $2Z_{\rm Ne}$ & $0.5Z_{\rm Ne}$ \\
    % & $\mu$m & [ &  & $10^3$ \msunyr & & & \hspace{1cm}] \\ 
    % \hline
    % \hline
    % O\,\textsc{vi} & 0.1037 & 0.55 & 0.05 & 0.27 & 1.60 & 0.63 & 0.35 \\
    % \lbrack Fe\,\textsc{viii}\rbrack & 5.4466 & 5 & 1 & 4 & 9 & 5 & 3 \\
    % % \lbrack Mg VII\rbrack & 5.5032 & - \\
    % % \lbrack Mg V\rbrack & 5.6099 & - \\
    % \lbrack Ne\,\textsc{vi}\rbrack & 7.6524 & 25 & 5 & 19 & 35 & 15 & 36 \\
    % \lbrack Mg\,\textsc{vii}\rbrack & 9.0090 & 6 & 1 & 5 & 11 & 7 & 4 \\
    % \lbrack Fe\,\textsc{vii}\rbrack & 9.5267 & 9 & 3 & 5 & 20 & 10 & 10 \\
    % \lbrack S\,\textsc{iv}\rbrack & 10.5105 & 31 & 3 & 33 & 39 & 30 & 34 \\
    % % \lbrack Ne II\rbrack & 12.8136 & - \\
    % \lbrack Ne\,\textsc{v}\rbrack & 14.3217 & 14 & 4 & 8 & 31 & 7 & 30 \\
    % \hline
    % \end{tabular}
    \begin{tablenotes}
        \item[1] The first four columns, from left to right, give (1) the ionization state, (2) the rest-frame vacuum wavelength of the emisison line, (3) the nominal cooling rate, and (4) the statistical uncertainty on the cooling rate. The remaining columns give the cooling rates, in units of 1000 \msunyr, when (5) the QSO bolometric luminosity is twice as bright, (6) the QSO bolometric luminosity is half as bright, (7) the hot phase neon abundance is twice as high, and (8) the hot phase neon abundance is half as high.
    \end{tablenotes}
\end{table*}

\subsection{\rev{Potential Alternative Sources for the Observed \nevi\, Emission}}
\label{sec:res_ionization}
\hfill\break
\rev{
In the main paper, we present our analysis of the data through a 2-component model where the 10$^{5.5}$ K coronal gas emission is due to gas that is cooling out of the ICM while also being heated by AGN photoionization, which we believe to be the model that represents the observations most accurately. In this section, we systematically go through other possible excitation sources for the observed coronal emission and determine the likelihood that they could also explain the observations.
}

\subsubsection{\rev{Shocks and Stellar Photoionization Heating}}
% -> maybe take out the AGN stuff from this section(?)

\rev{There are a number of mechanisms aside from AGN photoionization that could ionize the gas, including shocks and stellar photoionization}.  We calculate line ratios for a series of neon lines to avoid effects of metallicity and depletion. To calculate spaxel-by-spaxel line ratios, we first reproject our channel 4 line fluxes onto the channel 3 grid.  We then blur all of the line flux maps such that they have the same PSF FWHM as \nev. The resulting ionization diagram is shown in \rev{Supplementary Information Figure} \ref{fig:nevi_ionization}. We compare the observed ratios to a series of AGN photoionization and shock models. The AGN models are generated from [\citenum{2004ApJS..153...75G}] using a metallicity of $0.5Z_{\odot}$ and  density $n = 1000$ cm$^{-3}$. The grids show models with differing ionization parameters ($-4 < \log U < 0$) and power law slopes ($-2 < \alpha < -1.2$). The shock and precursor models are taken from the MAPPINGS III library\cite{2008ApJS..178...20A} using solar abundances and a magnetic field strength $B = 10$ $\mu$G. These grids run over velocity ($100 < v < 1000$ \kms) and density ($0.01 < n < 1000~{\rm cm}^{-3}$).

\begin{figure*}
    \centering
    \includegraphics[width=\textwidth]{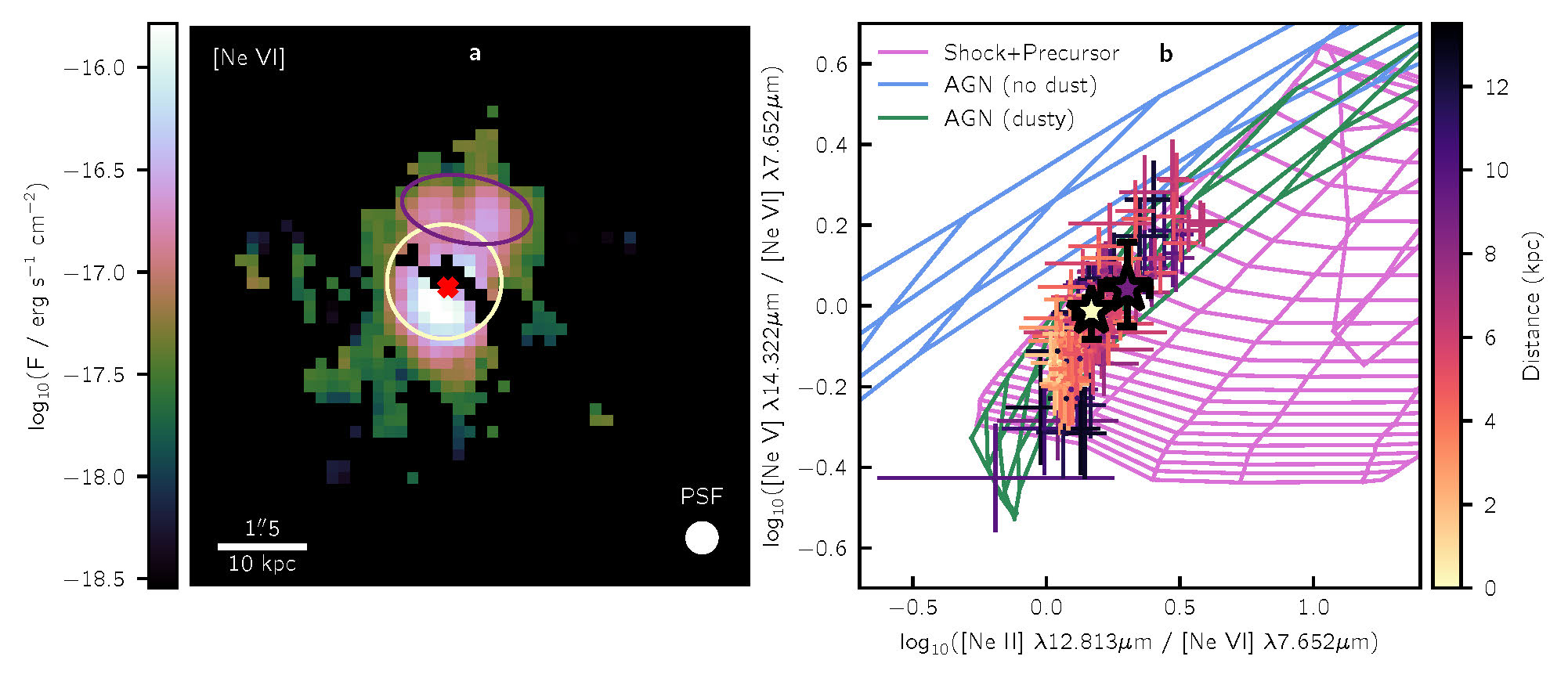}
    \caption{\textbf{An ionization diagram for the warm gas in the Phoenix Cluster using neon line ratios.} (a) The QSO-subtracted \nevi~flux is shown with apertures covering the nuclear region and northern extended region, shown in yellow and purple, respectively. Only spaxels with a detection of $S/N \geqslant 3$ are shown. (b) An ionization diagram using only neon lines. Model grids for AGN photoionization with and without dust, as well as shock ionization, are shown. Points correspond to individual spaxels and stars correspond to integrated ratios in the apertures shown in the left plot. Error bars represent 1$\sigma$ uncertainties. For spaxels within the nuclear aperture ($1''$ radius around the nucleus), we use non-QSO-subtracted fluxes, while for spaxels within the northern aperture we use the QSO-subtracted fluxes. The color scale of the data points indicates the distance from the nucleus, shown in the right colorbar.}
    \label{fig:nevi_ionization}
\end{figure*}

\rev{UV photons from young, hot stars do not get energetic enough to ionize coronal gas---even the hottest O-type stars have an insignificant flux of photons above 90 eV, the ionization threshold for coronal emission (\nevi\, has an ionization potential of 126.2 eV). However,} for the \neii~line, its lower IP \rev{of 21.6 eV} means that a significant fraction of its flux likely \rev{does come from stellar photoionization}. To address this, in the northern aperture we estimate the amount of \neii~flux that is due to young stars by measuring the [O II] $\lambda\lambda$3727,3729\angstrom~flux within the same aperture from the GMOS IFU data\cite{2014ApJ...784...18M} and taking a typical \neii/[O\,\textsc{ii}] ratio from the 8 Myr, 0.4$Z_\odot$ stellar photoionization models of [\citenum{2001ApJ...556..121K}]. This yields a \neii~flux of $\sim 8 \times 10^{-16}$ \ergscm. We subtract this flux from the observed \neii~flux within the aperture. The resulting ratios are all consistent with either dusty AGN photoionization or shocks.

Despite the consistent line ratios, shocks \rev{are} not sufficient to explain all of the observed \nevi\, flux. The shock models from the grid points closest to the observed data predict \nevi\, luminosities that are two orders of magnitude fainter than observed. They also require gas velocities exceeding 1000 \kms\,, which are not observed in the kinematics. Shocks, therefore, must contribute only a very small percentage of the total observed coronal emission. \rev{Nevertheless, the strong spatial correspondence between the axis of the northern radio jet emitted by the central AGN and a depression in the \nevi\, emission (see Supplementary Information $\S$\ref{sec:morph_compare}) suggests that, even if the jet is not important in driving the ionization of the coronal gas, it does play a significant role in shaping its morphology. The rapid time-variability of the AGN and its switching between radiative and mechanical feedback modes, which is thought to cause a disconnect between the observed X-ray and radio properties of this system\cite{2021A&A...646A..38T}, also implies that shocks from the jet may become an important source of ionization in times of lessened radiative AGN activity.}  

AGN photoionization \rev{is consistent with the observed line ratios and} is \rev{likely} the dominant source of coronal emission, but it cannot be the only source, as evidenced by the bump in the \nevi\, surface brightness and \nevi/\nev\, profiles (Figure \ref{fig:radialflux}).

\subsubsection{\rev{Cosmic Ray Heating}}

\rev{
Cosmic rays can also serve as a source of ionization and heating for the gas. We test whether cosmic ray heating has any substantial effect on the modeled coronal emission by running \textsc{Cloudy} models without the cosmic ray background ionization component. We find that the addition of cosmic rays has no effect on isobaric runs, and enhances integrated line emissivities ($\Gamma$) by 5--10\% on isochoric runs.  Overall, we find that it makes a negligible difference to the results.
}

\subsubsection{\rev{Turbulent Mixing and Conduction}} \label{sec:mixing_layers}

\rev{
It is also possible to create highly ionized emission if the hot atmosphere comes into contact with the cooler gas phases and the relative motions between the two gases at different densities and temperatures create Kelvin-Helmholtz and Rayleigh-Taylor instabilities that form a mixing layer at their interface. The mixing of the gases is driven by microturbulence, so the fact that we see enhanced line widths in the coronal gas (Extended Data Figure \ref{fig:nevi_kinematics}) supports a picture in which such turbulent mixing could be taking place.  The cool gas is heated and the hot gas is cooled collisionally until an intermediate temperature is reached\cite{1990MNRAS.244P..26B}. The cool gas contributes most of the mass to the mixed phase, whereas the hot gas contributes most of the energy\cite{1993ApJ...407...83S}.  The mixing layer then begins to cool radiatively back down to the cool phase, where it can be recycled and reheated (if there is still hot gas left over to mix with), starting the process over. In this model, radiative cooling from the hot phase to the mixed intermediate phase is severely suppressed, which may explain the missing soft X-ray emission lines if the intermediate gas temperature is below the peaks of their emissivity curves.
}

\rev{
We explore this possibility by running a series of \textsc{Cloudy} simulations with initial conditions representative of a mixed gas phase rather than the hot atmosphere.  For these simulations, the mixed phase is created at the interface of the ISM and the ICM. We do not believe that the hot phase mixes directly with the cold molecular gas, for 2 reasons: 1) the mixed phase is expected to settle non-radiatively to the geometric mean temperature between the hot and cold phases\cite{1990MNRAS.244P..26B}, which would imply that 10$^7$ K gas mixing with cold molecular gas at temperatures of 10--100 K creates a mixing layer at 1--3$\times 10^4$ K, far below the temperatures where coronal emission from \nevi\, and \ovi\, peak; 2) we do not observe a significant gradient in the molecular gas temperature along the filaments, which would be expected if it is partially mixing with hotter phases.  Therefore, we take the ISM temperature to be more representative of the warm ionized phase at 10$^4$ K. The ICM conditions are $n_H = 0.15$ cm$^{-3}$ and $kT = 6$ keV\cite{2015ApJ...811..111M}, representative of gas slightly further from the nucleus than the values used in our unmixed simulations.  We assume the gas phases mix in some mass proportion, ranging from ISM:ICM $=$ 2:1 to 7:1, making the mixed phase's initial density a multiple of 3--8$\times$ the hot phase density. The initial temperature of the mixed phase is chosen to keep it in pressure equilibrium with the hot phase\cite{2010ApJ...719..523K}. The metal abundances of the ISM use solar values\cite{2003ApJ...591.1220L}, with refractory elements depleted onto dust grains\cite{2009ApJ...700.1299J} with a depletion strength parameter $F_*$ ranging from 0.5--1.  Metal abundances of the ICM use the same values as our main simulations (measured from \textit{XMM} RGS spectra).  These abundances mix with the same ratios as the mass (2:1 to 7:1). We run these simulations with and without the incident AGN radiation, and under isobaric and isochoric cooling.  
}

\rev{
We find the model that best represents our observations to be the AGN-illuminated case with a mixing ratio of 4:1, a depletion strength $F_* = 0.6$, and composite isobaric/isochoric cooling with $x \simeq 0.12$ (see $\S$\ref{sec:mixedcooling}). Since this corresponds to denser gas than our main simulations, the cooling rates of all lines are driven marginally higher, and there is a smaller variance of the cooling rates between each line (as seen in the right panel of Figure \ref{fig:nev_nevi_ratio} in the main text), resulting in an average cooling rate of 15,000 $\pm$ 2,000 \msunyr\, over all the IR coronal lines.  The higher average causes the range of plausible cooling rates from our systematic uncertainty analysis to be expanded relative to the unmixed models, covering 7,000--36,000 \msunyr.  However, this analysis does not take into account the systematic uncertainties introduced by the new model parameters unique to the mixing scenario (the ISM:ICM mixing ratio and the depletion strength), so the range is likely underestimated. The line ratios of all the IR lines show better agreement with the observations than our unmixed simulations, which we believe to be primarily a result of the mixed phase metal abundances being more representative of the 10$^{5.5}$ K gas phase than the ICM abundances. A comparison of the line ratios is shown in Supplementary Information Figure \ref{fig:mixing_line_ratios}.  
}

\begin{figure}
    \rev{
    \centering
    \includegraphics[width=0.48\linewidth]{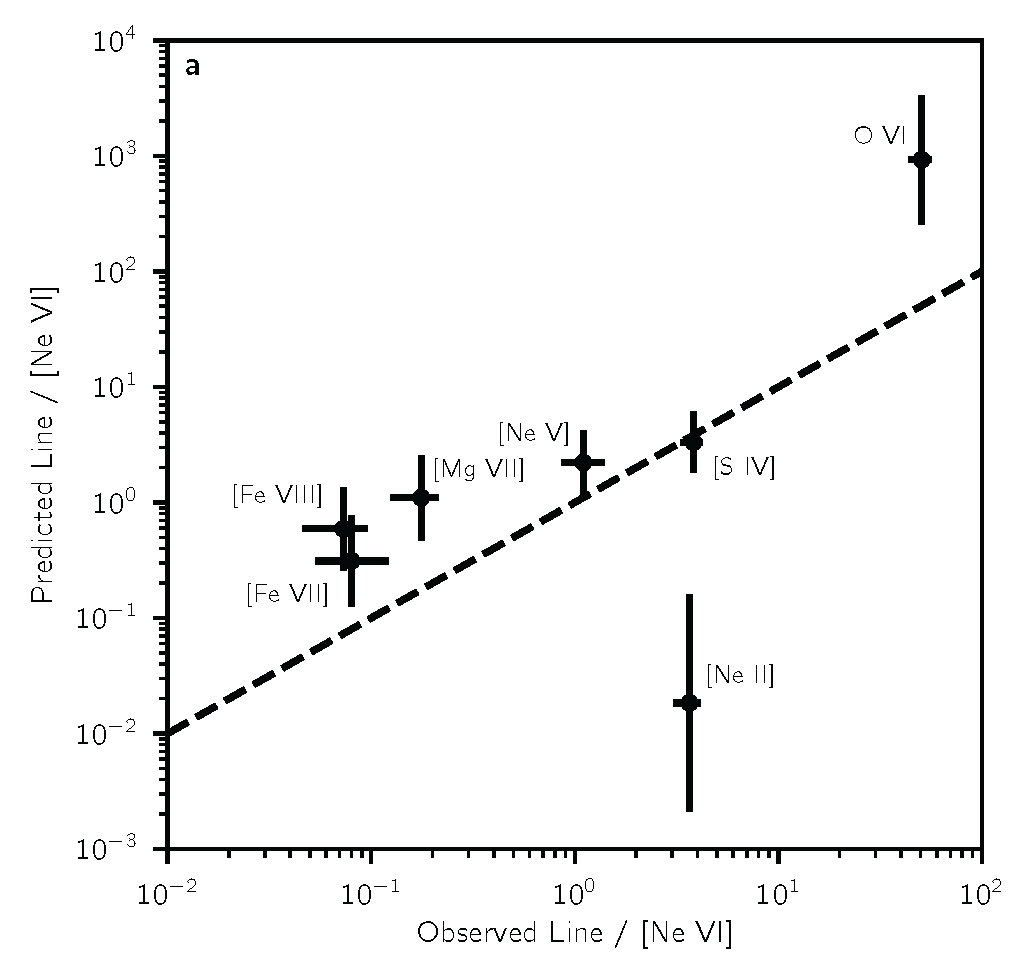}
    \includegraphics[width=0.48\linewidth]{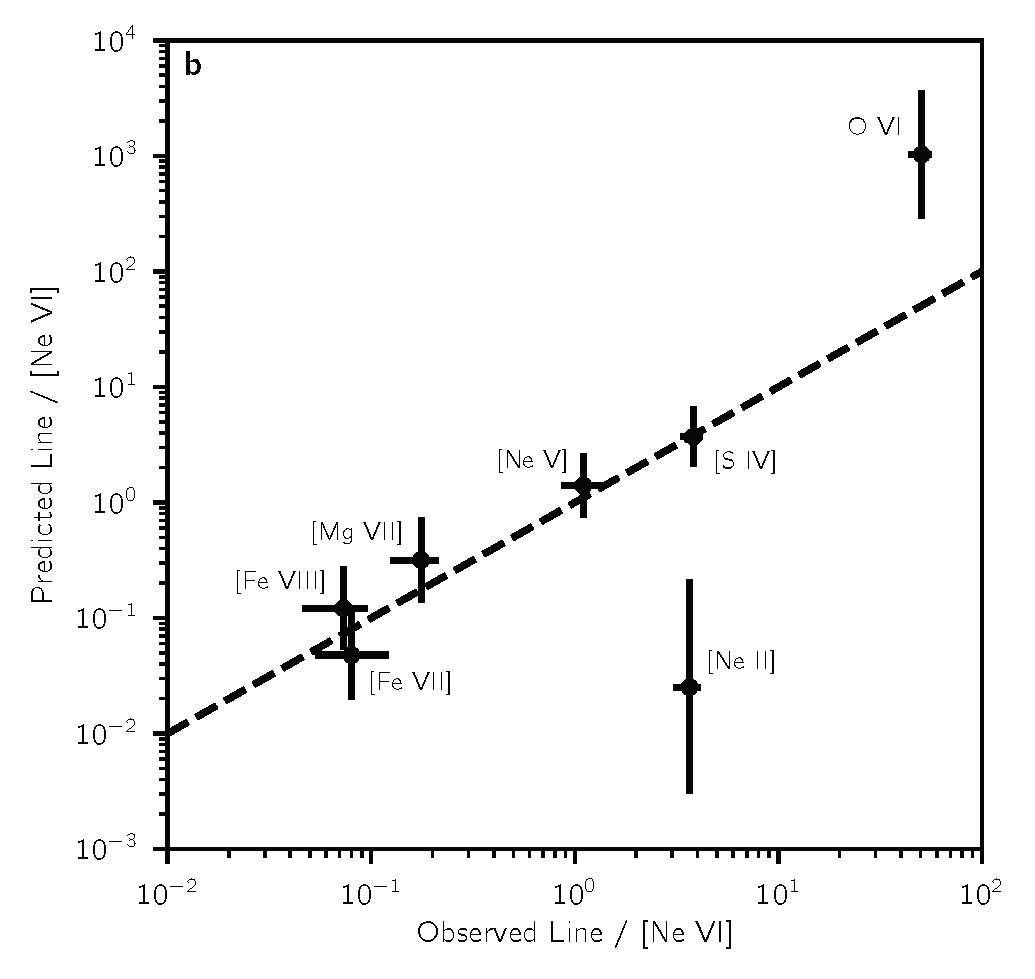}
    \caption{\textbf{Predicted line ratios from \textsc{Cloudy} simulations of cooling gas with illumination from an AGN and a combination of isobaric \& isochoric cooling.} The line ratios have been averaged over the northern aperture, using the observed line surface brightness at each radius as weights. The $x$-errors show the statistical 1$\sigma$ uncertainties in the observations, while the $y$-errors show the systematic uncertainties in the modeling. The dashed black line has a slope of unity, representing perfect agreement between the model and the data. Panel (a) shows the results for the unmixed simulation, while panel (b) shows the results for the mixing layer simulation.}
    \label{fig:mixing_line_ratios}
    }
\end{figure}

\begin{table}
    \centering
    \caption{\textbf{Buoyant bubble model starting parameters}}
    \label{tab:bubble}
    \includegraphics[width=0.5\textwidth]{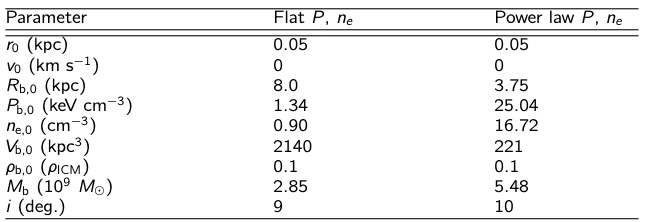}
    % \begin{tabular}{lll}
    % \hline
    % {Parameter} & {Flat $P$, $n_e$} & {Power law $P$, $n_e$} \\
    % \hline
    % \hline
    % $r_0$ (kpc) & 0.05 & 0.05 \\
    % $v_0$ (\kms) & 0 & 0 \\
    % $R_{\rm b,0}$ (kpc) & 8.0 & 3.75 \\
    % $P_{\rm b,0}$ (keV cm$^{-3}$) & 1.34 & 25.04 \\
    % $n_{\rm e,0}$ (cm$^{-3}$) & 0.90 & 16.72 \\
    % $V_{\rm b,0}$ (kpc$^3$) & 2140 & 221 \\
    % $\rho_{\rm b,0}$ ($\rho_{\rm ICM}$) & 0.1 & 0.1 \\
    % $M_{\rm b}$ ($10^{9}$ \msun) & 2.85 & 5.48 \\
    % $i$ (deg.) & 9 & 10 \\
    % \hline
    % \end{tabular}
\end{table}

\clearpage

%%%%%%%%%%%%%%%%%%%%%%%%%%%%%%%% END NOTES %%%%%%%%%%%%%%%%%%%%%%%%%%%

\begin{addendum}

\item[Acknowledgements]
This work is based on observations with the NASA/ESA/CSA James Webb Space Telescope obtained from the Data Archive at the Space Telescope Science Institute, which is operated by the Association of Universities for Research in Astronomy, Incorporated, under NASA contract NAS 5-03127. Support for Program number JWST-GO-02439.001-A was provided through a grant from the STScI under NASA contract NAS 5-03127. 

This work is also based in part on observations made with the NASA/ESA Hubble Space Telescope obtained from the Space Telescope Science Institute, which is operated by the Association of Universities for Research in Astronomy, Inc., under NASA contract NAS 5–26555. These observations are associated with program GO15315. 

Support for this work was also provided by NASA through \textit{Chandra} Award Number GO7-18124 issued by \textit{Chandra}, which is operated by the Smithsonian Astrophysical Observatory for and on behalf of the National Aeronautics Space Administration under contract NAS8-03060. 

M. Reefe acknowledges support by the National Science Foundation Graduate Research Fellowship under Grant No. 2141064. 

M. Gaspari acknowledges support from the ERC Consolidator Grant \textit{BlackHoleWeather} (101086804).

M. Chatzikos also acknowledges support from NSF (1910687), and NASA (19-ATP19-0188, 22-ADAP22-0139).

H. Russell acknowledges an Anne McLaren Fellowship from the University of Nottingham

We would like to thank the members of the MIRI/MRS instrument team, particularly D. Law, for providing advice and guidance in reducing and cleaning MIRI/MRS data. 

\item[Author Contributions]
M. Reefe reduced the MIRI/MRS data, wrote and ran the spectral fitting and PSF decomposition code, did the main analysis of the ionization, kinematics, and cooling, and wrote the paper. M. McDonald provided substantial guidance in the main analyses, in the interpretation of the results, and in writing the paper. He also wrote the original proposal for MIRI/MRS data. M. Chatzikos performed the \textsc{Cloudy} simulations and wrote the description of these processes in the methods section. J. Seebeck performed a sanity check on our data reduction and spectral modeling by running an independent pipeline (Q3Dfit). K. Sharon performed strong lensing models to estimate the gravitational potential, which we used in our kinematic modeling. R. Mushotzky, S. Veilleux, S. Allen, M. Bayliss, M. Calzadilla, R. Canning, B. Floyd, M. Gaspari, J. Hlavacek-Larrondo, B. McNamara, H. Russell, and T. Somboonpanyakul provided feedback on the paper drafts.

\item[Author Information]
The authors declare that they have no competing financial interests. \rev{Supplementary Information is available for this paper.} Correspondence and requests for materials should be addressed to M. Reefe (mreefe@mit.edu). Reprints and permissions information is available at www.nature.com/reprints.

\item[Data Availability]
\textit{JWST} program ID 2439 data \rev{is} publicly available through the Space Telescope Science Institute's (STSci) Mikulski Archive for Space Telescopes (MAST). The supplementary data used in our analysis from \textit{HST} is also available on MAST, and the \textit{Chandra} data is available from the Chandra Data Archive.

\item[Code Availability]
The \textit{JWST}, \textit{HST}, and \textit{Chandra} data were reduced using the publicly available reduction pipeline codes provided by STSci and the Chandra X-ray Center. The LOKI code used in our analysis is available on GitHub: \href{https://github.com/Michael-Reefe/Loki.jl}{https://github.com/Michael-Reefe/Loki.jl}, \rev{and \textsc{Cloudy} is available on GitLab: \href{https://gitlab.nublado.org/cloudy/cloudy}{https://gitlab.nublado.org/cloudy/cloudy}. We also provide our customized driver scripts for the data reduction and analysis at \href{https://github.com/Michael-Reefe/Reefe2024_code_supplements}{https://github.com/Michael-Reefe/Reefe2024\_code\_supplements}.}

\end{addendum}

\end{document}